\documentclass[12pt]{iopart}
\pdfoutput=1 
\usepackage{graphicx}
\usepackage{amsfonts}
\usepackage{amstext}
\usepackage{amssymb}
\usepackage{cite}
\usepackage{color}
\usepackage{epstopdf}

\def\R{\mathbb R}

\def\ve{\varepsilon}

\def\K{{\mathcal K}}

\begin{document}

\title{Mean perimeter and mean area of the convex hull over planar random walks}

\author{Denis~S.~Grebenkov}
 \ead{denis.grebenkov@polytechnique.edu}
\address{
Laboratoire de Physique de la Mati\`{e}re Condens\'{e}e (UMR 7643), \\ 
CNRS -- Ecole Polytechnique, 91128 Palaiseau, France}

\address{Interdisciplinary Scientific Center Poncelet (ISCP),%
\footnote{International Joint Research Unit -- UMI 2615 CNRS/ IUM/ IITP RAS/ Steklov MI RAS/ Skoltech/ HSE, Moscow, Russian Federation} \\
Bolshoy Vlasyevskiy Pereulok 11, 119002 Moscow, Russia}

\author{Yann Lanoisel\'ee}
\address{
Laboratoire de Physique de la Mati\`{e}re Condens\'{e}e (UMR 7643), \\ 
CNRS -- Ecole Polytechnique, 91128 Palaiseau, France}

\author{Satya~N.~Majumdar}
\address{
Laboratoire de Physique Th\'eorique et Mod\`eles Statistiques (UMR 8626 du CNRS), \\
Universit\'e de Paris-Sud, B\^at. 100, 91405 Orsay Cedex, France}

\begin{abstract}
We investigate the geometric properties of the convex hull over $n$
successive positions of a planar random walk, with a symmetric
continuous jump distribution.  We derive the large $n$ asymptotic
behavior of the mean perimeter.  In addition, we compute the mean area
for the particular case of isotropic Gaussian jumps.  While the
leading terms of these asymptotics are universal, the subleading
(correction) terms depend on finer details of the jump distribution
and describe a ``finite size effect'' of discrete-time jump processes,
allowing one to accurately compute the mean perimeter and the mean
area even for small $n$, as verified by Monte Carlo simulations.  This
is particularly valuable for applications dealing with discrete-time
jumps processes and ranging from the statistical analysis of
single-particle tracking experiments in microbiology to home range
estimations in ecology.
\end{abstract}





\date{\today}

\maketitle

\section{Introduction}

Consider a set of $n$ points with position vectors $\{\vec r_1, \vec
r_2,\ldots, \vec r_n\}$ in a $d$-dimensional space.  The most natural
and perhaps the simplest way to characterize the {\it shape} of this
set of points is by drawing the convex hull around this set: a convex
hull is the unique minimal convex polytope that encloses all the
points.  This unique polytope is convex since the line segment joining
any two points on the surface of the polytope is fully contained
within the polytope.  Properties of such convex polytopes have been
widely studied in mathematics, computer science (image processing and
patter recognition) and in the physics of crystallography (the Wulff
construction).  In two dimension, where the convex hull is a polygon,
there are many other applications, most notably in ecology where the
home range of animals or the spread of an epidemics are typically
estimated by convex hulls. For a review on the history and
applications of convex hulls, see Ref.~\cite{Majumdar2010a}.

When the points $\{\vec r_1, \vec r_2,\ldots, \vec r_n\}$ are drawn
randomly from a joint distribution $P(\vec r_1, \vec r_2,\ldots, \vec
r_n)$, the associated convex hull also becomes random and
characterizing its statistical properties is a challenging problem,
since the convex hull is a highly nontrivial functional of the random
variables $\{\vec r_1, \vec r_2,\ldots, \vec r_n\}$.  For instance,
what can one say about the statistics of the surface area $S_d$ or the
volume $V_d$ of the convex hull, for a given joint distribution
$P(\vec r_1, \vec r_2,\ldots, \vec r_n)$?  Even finding the mean
surface area $\langle S_d\rangle$ or the mean volume $\langle
V_d\rangle$, for arbitrary joint distribution, is a formidably
difficult problem.  In the special case when the points are
independent and identically distributed, i.e., when the joint
distribution factorizes as $P(\vec r_1, \vec r_2,\ldots, \vec r_n)=
\prod_{k=1}^n P(\vec r_k)$ (with $P(\vec r_k)$ representing the
marginal distribution), several results on the statistics of the
surface and volume of the convex hull are known
(see~\cite{Majumdar2010a} for a historical review).  However, for {\it
correlated} points where the joint distribution does not factorize,
very few results are available.

The simplest example of a set of correlated points corresponds to the
case of a random walk in $d$-dimensional continuous space, where $\vec
r_k$ represents the position of the walker at step $k$, starting at
the origin at step $0$.  The position evolves via the Markov rule,
$\vec r_k= \vec r_{k-1}+ \vec \eta_k$, where $\vec \eta_k$ represents
the jump at step $k$, and one assumes that $\vec \eta_k$ are
independent and identically distributed random variables, each drawn
from some prescribed distribution $p(\vec \eta_k)$. The walk evolves
up to $n$ steps generating the vertices $\{\vec r_1, \vec r_2,\ldots,
\vec r_n\}$ of its trajectory.  There is a unique convex hull for each
sample of this trajectory and what can one say about the mean surface
area or the mean volume of this convex hull, given the jump
distribution $p(\vec \eta_k)$?  This is the basic problem that we
address in this paper.  We show that at least for $d=2$ (planar random
walks), it is possible to obtain precise {\it explicit} results for
all $n$ for the mean perimeter and the mean area of the convex hull of
the walk, for a large class of jump distributions $p(\vec \eta)$,
including in particular L\'evy flights where the jump distribution has
a fat tail.  We also obtain similar results for the mean area of the
convex hull but under additional assumptions on the jump distribution.

This problem concerning the convex hull of a random walk becomes
somewhat simpler in the special case of the Brownian limit, where
several results are known.  Consider, for example a jump distribution
$p(\vec \eta_k)$ with zero mean and a finite variance $\sigma^2$.  In
this case, the walk converges in the large $n$ limit to the Brownian
motion.  In other words, one can consider the continuous-time limit,
as $\sigma^2\to 0$ and $n\to \infty$ with $n\sigma^2= 2\, D\, t$ being
fixed (here $D$ is called the diffusion constant and $t$ is the
duration of the walk).  In this Brownian limit and for $d=2$, the mean
perimeter and the mean area have been known exactly for a while.
Tak\'acs~\cite{Takacs1980} computed the mean perimeter
\begin{equation}
\langle S_2\rangle =  \sqrt{16 \pi\, D\, t}\, ,
\label{eq:perim_BM}
\end{equation}
while El Bachir~\cite{ElBachir83} and Letac~\cite{Letac93} 
computed the mean area
\begin{equation}
\langle V_2\rangle = \pi\, D\, t\, .
\label{eq:area_BM} 
\end{equation}
For a planar Brownian bridge of duration $t$ (where the walker returns
to the origin after time $t$), the mean perimeter $\langle
S_2\rangle_{\rm bridge} = \sqrt{\pi^3\, D\, t}$ was computed by
Goldman~\cite{Goldman96}, while the mean area $\langle V_2\rangle_{\rm
bridge}=(2\pi/3)\, D\, t$ was computed relatively recently by
Randon-Furling {\it et al.}~\cite{Randon09}.  An interesting extension
of this problem in $d=2$ is to the case of $N$ independent planar
Brownian motions (or Brownian bridges)~\cite{Randon09,Majumdar2010a}.
This is relevant in the context of the home range of animals, where
$N$ represents the size of an animal population and the trajectory of
each animal is approximated by a Brownian motion during their foraging
period.  For a fixed population size $N$, the mean perimeter and the
mean area of the convex hull was computed exactly: $\langle S_2\rangle
= \alpha_N \, \sqrt{D\,t}$ and $\langle V_2 \rangle = \beta_N\, D\,
t$, where the prefactors $\alpha_N$ and $\beta_N$ were found to have
nontrivial $N$ dependence \cite{Randon09,Majumdar2010a}.  For $d>2$,
very few exact results are known for this problem.  For a single
$(N=1)$ Brownian motion, the mean surface area and the mean volume of
the convex hull was recently computed by Eldan~\cite{Eldan2014}:
$\langle S_d \rangle=\frac{2(4\pi D t)^{(d-1)/2}}{\Gamma(d)}$ and
$\langle V_d\rangle= \frac{(\pi D t)^{d/2}}{\Gamma(d/2+1)^2}$ (see
also \cite{Kabluchko16b} for another derivation and extension to
Brownian bridges).  However, for $N>1$ and $d>2$, no exact result is
available.  Finally, going beyond the mean surface and the mean
volume, very few results are known for higher moments (see the review
\cite{Majumdar2010a} for results on variance) or even the full
distribution of the surface or the volume of the convex hull of
Brownian motion (see Refs.~\cite{Wade15,Wade15b} for a recent
discussion on the distribution of the perimeter in $d=2$ and $N=1$).
Very recently, the full distribution (including the large deviation
tails) of the perimeter and the area of $N\ge 1$ planar Brownian
motions were calculated numerically~\cite{Claussen15,Dwenter16}.  Some
rigorous results on the convex hulls of L\'evy processes were recently
derived \cite{Molchanov12,Molchanov16}.

If one is interested only in the mean area or the mean volume of the
convex hull of a generic stochastic process (not necessarily just a
random walk), a particular simplification occurs in $d=2$ (planar
case) where several analytical results can be derived by adapting
Cauchy's formula~\cite{Cauchy1832,Santalo} for arbitrary closed convex
curves in $d=2$.  Indeed by employing Cauchy's formula for every
realization of a random planar convex hull, it was shown in
Refs.~\cite{Randon09,Majumdar2010a} that the problem of computing the
mean perimeter and the mean area of an {\it arbitrary} two dimensional
stochastic process (can in general be non-Markovian and in
discrete-time) can be mapped to computing the extremal statistics
associated with the one dimensional component of the process (see
Section 3 for the precise mapping).  This mapping was introduced
originally in~\cite{Randon09} to compute $\langle S_2\rangle$ and
$\langle V_2\rangle$ exactly for $N\ge 1$ planar Brownian motions.
Since then, it has been used for a number of continuous-time planar
processes: random acceleration process~\cite{Reymbaut11}, branching
Brownian motion with applications to animal epidemic
outbreak~\cite{Dumonteil13}, anomalous diffusion
processes~\cite{Lukovic13} and also to a single Brownian motion
confined to a half-space~\cite{Chupeau2015a}.

The objective of this paper is to go beyond the continuous-time limit
and obtain results for the convex hull of a discrete-time planar
random walk of $n$ steps (with $n$ large but finite) with arbitrary
jump distribution, including for instance L\'evy flights.  Indeed, in
any realistic experiment or simulation, the points of the trajectory
are always discrete.  For example, recently proposed local convex hull
estimators \cite{Lanoiselee17} are based on a relatively small number
of points, where we can not apply the Brownian limiting results
reviewed above.  The first rigorous result for a two-dimensional
discrete random walk, modeled as a sum of independent random variables
in the complex plane, was derived for the mean perimeter of the convex
hull by Spitzer and Widom \cite{Spitzer1961},
\begin{equation} \label{eq:Spitzer}
\langle L_n\rangle = 2\sum_{k=1}^n\frac{\langle \vert x_k + i y_k \vert \rangle}{k}
\end{equation}
(here $x_k+iy_k$ is the complex-valued position of the walker after
$k$ steps).  Although the formula (\ref{eq:Spitzer}) looks deceptively
simple, an {\it explicit} computation of the mean $\langle L_n
\rangle$ is difficult using Eq. (\ref{eq:Spitzer}), in particular
its behavior for large but finite $n$.  For the case of zero mean and
finite variance jump distributions, the leading $n^{\frac12}$ term in
$\langle L_n\rangle$ was identified in \cite{Spitzer1961}, but the
relevant subleading terms were not known, to the best of our
knowledge.  For other results on the statistics of $L_n$, see
Ref.~\cite{Snyder93}.  The Spitzer-Widom formula (\ref{eq:Spitzer})
was extended to generic $d$-dimensional random walks in
Ref. \cite{Vysotsky15}, in which exact combinatorial expressions for
the expected surface area and the expected volume of the convex hull
were derived.  However, these expressions are not suitable for the
asymptotic analysis at large $n$.  Several other geometrical
properties of the convex hull of random walks are known: for example,
the exact formula for the mean number of facets of the convex polytope
of a $d$-dimensional random walk, for $d=2$~\cite{Baxter61} and $d >
2$~\cite{Kabluchko16,Randon17}.  But in this paper, we will restrict
ourselves only to the mean perimeter $\langle L_n\rangle$ and the mean
area $\langle A_n\rangle$ of a planar random walk of $n$ steps and our
main goal is to derive explicitly not only the leading term in
$\langle L_n\rangle$ and $\langle A_n\rangle$ for large $n$, but also
the subleading terms.

Our strategy is to adapt the mapping between the convex hull of a
$2$-d process and the extreme statistics of the $1$-d component
process, mentioned above, to the case of a single discrete-time planar
random walk with generic jump distributions.  Using this strategy, we
are able to compute explicitly the leading and subleading terms of the
mean perimeter of the convex hull of a planar random walk of $n$ steps
with arbitrary symmetric continuous jump distributions for large but
finite $n$.  The mean area is also computed but only for the
particular case of isotropic Gaussian jumps.  The rest of the paper is
organized as follows.  Section \ref{sec:outline} outlines the class of
considered planar random walks and the main results.  In
Sec. \ref{sec:theory}, we explain the derivation steps.  In
Sec. \ref{sec:simu}, several explicit examples are presented and used
to illustrate the accuracy of the derived asymptotic relations by
comparison with Monte Carlo simulations.  In
Sec. \ref{sec:discussion}, we discuss the main results, their
applications, and conclusions.  \ref{sec:Aderivation} and
\ref{sec:Aexamples} contain some technical details of the derivation
and exactly solvable examples, respectively.

\section{The model and the main results}
\label{sec:outline}

\begin{figure}  
\begin{center}
\includegraphics[width=80mm]{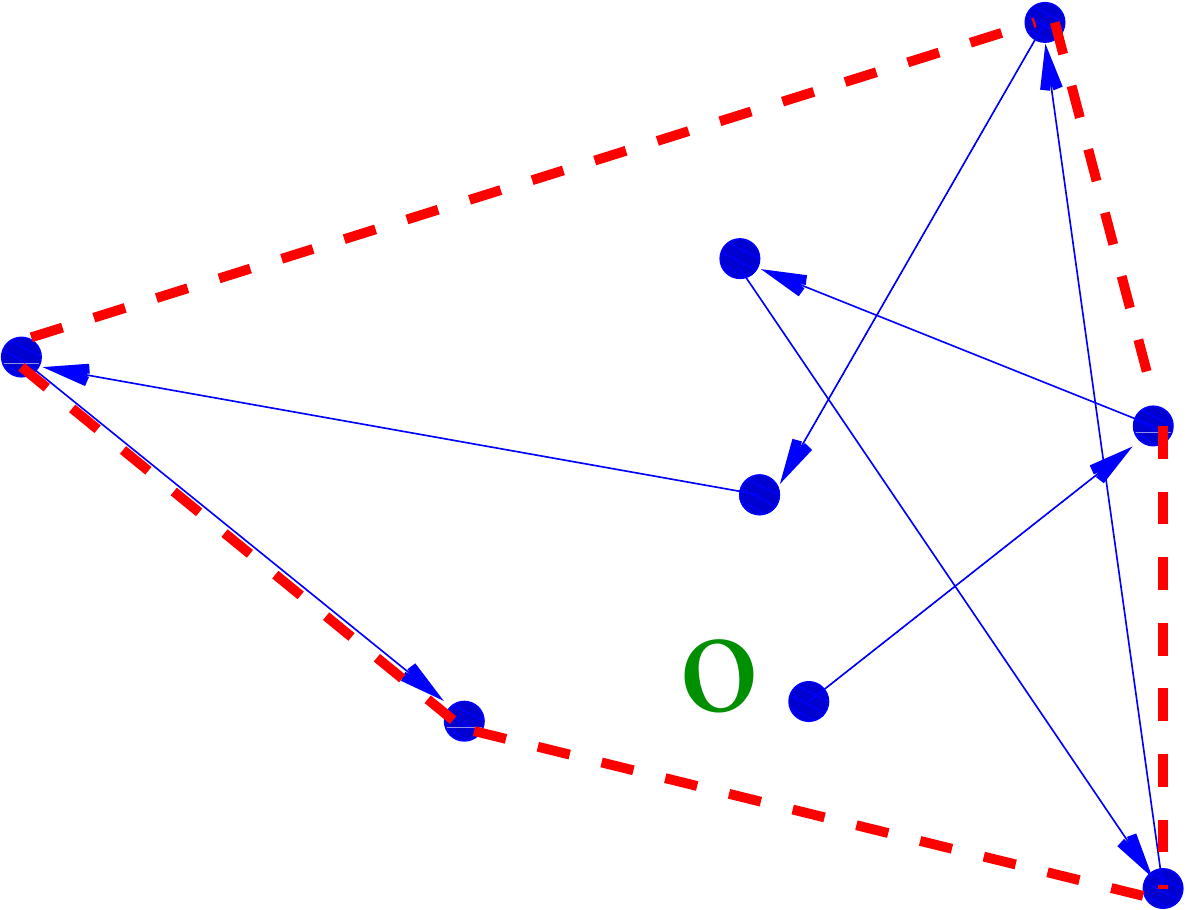}
\end{center}  
\caption{
Illustration of the convex hull of a $7$-stepped planar random
walk. The walk starts at the origin $O$ and makes independent jumps at
each step (shown by arrows). after $7$ steps, the convex hull (shown
by dashed red lines) is constructed around the points of the
trajectory.}
\label{fig:conhull_rw}
\end{figure}

We consider a discrete-time random walker in the plane whose jumps are
random, independent, and identically distributed.  Starting from the
origin, the walker produces a sequence of $(n+1)$ points $\{(x_0,y_0),
(x_1,y_1), \ldots, (x_n,y_n)\}\subset \R^2$ after $n$ jumps such that
\begin{equation}  \label{eq:RW_def}
\fl 
(x_0,y_0) = (0,0), \qquad (x_k,y_k) = (x_{k-1}, y_{k-1}) + (\eta_k^x,\eta_k^y)  \quad (k=1,2,\ldots,n),
\end{equation}
where the jumps $\vec \eta_k = (\eta_k^x,\eta_k^y)$ at the $k$-th step
are independent from step to step, and at each step they are drawn
from a prescribed joint probability density function (PDF) $p(x,y)$,
i.e.,
\begin{equation}
\P\{\eta_k^x \in (x, x+dx), \eta_k^y \in (y, y+dy)\} = p(x,y)\,  dx\,dy\, .
\label{noisepdf}
\end{equation}
We emphasize that the starting point $(x_0,y_0)$ is not random and for
convenience, we choose $(x_0= y_0=0)$ to be the origin.  The convex
hull constructed over these $(n+1)$ points is the minimal convex
polygon that encloses all these points (see Fig. \ref{fig:conhull_rw}
for an illustration).  We are interested in the perimeter $L_n$ and
the area $A_n$ of the convex hull which are random variables given
that the points are generated as successive positions of a planar
random walk.  We aim at computing {\it exactly} the leading and
subleading terms of the mean perimeter, $\langle L_n\rangle$, and the
mean area, $\langle A_n\rangle$, of the convex hull for large $n$.

As explained in Sec. \ref{sec:theory}, our computation relies on two
key results: (i) Cauchy's formula for the perimeter and the area of a
closed convex curve, that allows one to reduce the original planar
problem to the analysis of one-dimensional projections, and (ii) the
Pollaczek-Spitzer formula describing the distribution of the maximum
of partial sums of independent symmetric continuously distributed
random variables~\cite{Pollaczek52,Spitzer56}.  To use the
Pollaczek-Spitzer formula, we need thus to assume that the joint
probability density $p(x,y)$ is continuous and centrally symmetric:
\begin{equation}  \label{eq:p_symm}
p(-x,-y) = p(x,y) .
\end{equation}
In particular, our results will not be applicable to a classical
random walk on the square lattice because its distribution is not
continuous.  In the following, we outline the main results that will
be derived in Sec. \ref{sec:theory}.

The mean perimeter $\langle L_n\rangle$ is computed for a very general
class of symmetric continuous jump distributions.  Writing the Fourier
transform of $p(x,y)$ as
\begin{equation}  \label{eq:hatrho_xi_general}
\hat{\rho}_\theta(k) = \int\limits_{-\infty}^\infty dx 
\int\limits_{-\infty}^\infty dy \, p(x,y) \, e^{ik (x\cos\theta + y \sin\theta)}  ,
\end{equation}
one can characterize the behavior of the mean perimeter according to
the asymptotic properties of $\hat{\rho}_\theta(k)$ as $k\to 0$.  We
assume a general expansion
\begin{equation}  \label{eq:hatrho_mu}
\hat{\rho}_\theta(k) \simeq 1 - |a_\theta k|^\mu + o(|k|^\mu) \qquad (k\to 0) ,
\end{equation}
with the scaling exponent $0 < \mu \leq 2$, and a scale $a_\theta >
0$.  When $0 < \mu \leq 1$, the mean perimeter of the convex hull is
infinite.  We therefore focus on the case $1 < \mu \leq 2$.

First, we derive an exact formula for the generating function of
$\langle L_n\rangle$ which is valid for any $1 < \mu \leq 2$.
Extracting the asymptotic large $n$ behavior of $\langle L_n\rangle$
from this general formula is, however, nontrivial.  We distinguish the
following cases.

\begin{enumerate}
\item 
When the jump variance is finite ($\mu = 2$), the mean perimeter
is shown to behave as
\begin{equation}  \label{eq:Lmean_asympt0}
\langle L_n\rangle \simeq C_0 \, n^{\frac12} + C_1 + o(1) \qquad (n \gg 1),
\end{equation}
with
\begin{equation}  \label{eq:CLn}
C_0 = \frac{\sqrt{2}}{\sqrt{\pi}} \int\limits_0^{2\pi} d\theta \, \sigma_\theta ,  \qquad
C_1 = \int\limits_0^{2\pi} d\theta \, \sigma_\theta \, \gamma_\theta ,  \quad
\end{equation}
where $\sigma_\theta$ and $\gamma_\theta$ are given by
\begin{eqnarray}  
\label{eq:sigma_theta}
\sigma_\theta^2 &=& - \lim\limits_{k\to 0} \frac{\partial^2 \hat{\rho}_\theta(k)}{\partial k^2} = 
\langle (\eta^x \cos \theta + \eta^y \sin\theta)^2 \rangle = \frac{a_\theta^2}{2} \,, \\
\label{eq:gamma}
\gamma_\theta &=& \frac{1}{\pi \sqrt{2}} \int\limits_0^\infty \frac{dk}{k^2} 
\ln \left(\frac{1 - \hat{\rho}_\theta\bigl(\sqrt{2}\,k/\sigma_\theta\bigr)}{k^2}\right)  .
\end{eqnarray}
If in addition the fourth-order moment of the jump distribution is
finite, one gains the second subleading term,
\begin{equation}  \label{eq:Lmean_asympt}
\langle L_n\rangle \simeq C_0 \, n^{\frac12} + C_1 + C_2 \, n^{-\frac12} + o(n^{-\frac12}) \qquad (n \gg 1),
\end{equation}
with
\begin{equation}
C_2 = \frac{C_0}{8} + \frac{\sqrt{2}}{24\sqrt{\pi}} \int\limits_0^{2\pi} d\theta \, \sigma_\theta \, \K_\theta 
\end{equation}
and
\begin{equation}
\label{eq:a4}
\K_\theta = \frac{1}{\sigma_\theta^4} \lim\limits_{k\to 0} \frac{\partial^4 \hat{\rho}_\theta(k)}{\partial k^4} 
= \frac{\langle (\eta^x \cos \theta + \eta^y \sin\theta)^4 \rangle}{\langle (\eta^x \cos \theta + \eta^y \sin\theta)^2 \rangle^2} \,.
\end{equation}
Higher-order corrections can also be derived under further moments
assumptions.  Note that the integral expression for the coefficient
$C_0$ in front of the leading term $n^{1/2}$ first appeared in
\cite{Spitzer1961}.  In Sec. \ref{sec:simu}, we will show that the
asymptotic formula (\ref{eq:Lmean_asympt}) is very accurate even for
small $n$.

\item
When the jump variance is infinite (i.e., $1 < \mu < 2$), one needs to
consider the subleading term in the small $k$ asymptotics of
$\hat{\rho}_\theta(k)$:
\begin{equation}  \label{eq:hatrho_nu}
\hat{\rho}_\theta(k) \simeq 1 - |a_\theta k|^\mu + b_\theta |k|^\nu + o(|k|^\nu) \qquad (k\to 0) ,
\end{equation}
with the subleading exponent $\nu > \mu$ and a coefficient $b_\theta$.
Depending on the subleading exponent $\nu$, we distinguish two cases:

\subitem 
(1) if $\mu < \nu < \mu+1$, one has
\begin{equation}  \label{eq:Ln_Levy1}
\langle L_n\rangle \simeq C_0 \, n^{1/\mu} + C_1 \, n^{1-(\nu-1)/\mu} + o(n^{1-(\nu-1)/\mu})  \qquad (n\gg 1),
\end{equation}
with
\begin{eqnarray}  \label{eq:C0_Levy1}
C_0 &=& \frac{\mu \,\Gamma(1- 1/\mu)}{\pi} \int\limits_0^{2\pi} d\theta \, a_\theta   , \\
\label{eq:C1_Levy1}
C_1 &=& - \frac{\Gamma((\nu-1)/\mu)}{\pi (\mu+1-\nu)} \int\limits_0^{2\pi} d\theta \, a_\theta^{1-\nu} \, b_\theta .
\end{eqnarray}
Note that the coefficient $C_0$ also appears in the mean perimeter of
the convex hull of continuous-time symmetric stable processes
\cite{Molchanov12}.

\subitem
(2) if $\nu > \mu+1$, one has
\begin{equation}  \label{eq:Ln_Levy2}
\langle L_n\rangle \simeq C_0 \, n^{1/\mu} + C_1 + o(1)   \qquad (n\gg 1),
\end{equation}
with $C_0$ from Eq. (\ref{eq:C0_Levy1}) and 
\begin{equation}  \label{eq:C1_Levy2}
C_1 = \int\limits_0^{2\pi} d\theta \, \gamma_\theta,
\end{equation}
where
\begin{equation}  \label{eq:gamma_Levy2}
\gamma_\theta = \frac{1}{\pi} \int\limits_0^{\infty} \frac{dk}{k^2} \ln \left(\frac{1- \hat{\rho}_\theta(k)}{(ak)^\mu}\right) .
\end{equation}
For instance, for a L\'evy symmetric alpha-stable distribution with
$\hat{\rho}(k) = \exp(-|ak|^{\mu})$, one gets \cite{Comtet05}
\begin{equation}  \label{eq:gamma_Levystable}
\gamma = a \, \frac{\zeta(1/\mu)}{(2\pi)^{1/\mu} \sin(\pi/(2\mu))} \, ,
\end{equation}
where $\zeta(z)$ is the Riemann zeta function.
\end{enumerate}
The obtained results are indeed very general.

In turn, our method of computation of the mean area requires two
additional strong assumptions: (a) the independence of the jumps along
$x$ and $y$ coordinates, i.e., $p(x,y)= p(x) p(y)$ and (b) the
isotropy of the jump PDF, i.e., $p(x,y)$ should depend only on the
distance $r=\sqrt{x^2+y^2}$ but not on the direction of the jump.
According to Porter-Rosenzweig theorem~\cite{Porter60}, only the
Gaussian jump distribution with identical variance $\sigma^2$ along
$x$ and $y$ directions, i.e., $p(x,y)=
\frac{1}{2\pi \sigma^2}\,\exp[-(x^2+y^2)/{2\sigma^2}]$, satisfies
both properties (a) and (b).  Our result for the mean area is thus
only valid for this Gaussian distribution:
\begin{equation}   \label{eq:Amean_asympt}
\sigma^{-2} \langle A_n \rangle = \frac{\pi}{2} n + \gamma \sqrt{8\pi} \, n^{\frac12} + \pi (\K/12 + \gamma^2) + o(1) ,
\end{equation}
with $\K = 3$ and 
\begin{equation}  \label{eq:gamma_Gauss}
\gamma = \frac{\zeta(1/2)}{\sqrt{2\pi}} = - 0.58259 \ldots
\end{equation}
Recently, an exact formula the mean area of the convex hull of a
Gaussian random walk was derived \cite{Kabluchko16b}.  In the
isotropic case, the formula reads
\begin{equation}  \label{eq:sum_Gauss}
\sigma^{-2} \langle A_n \rangle = \frac12 \sum\limits_{i=1}^n \sum\limits_{j=1}^{n-i} \frac{1}{\sqrt{ij}} \,.
\end{equation}
While the result in Eq. (\ref{eq:sum_Gauss}) is very useful for finite
$n$, deriving the large $n$ asymptotics of this double sum (including
up to two subleading terms as in Eq. (\ref{eq:Amean_asympt})) seems
somewhat complicated.  Our method, in contrast, gives a more direct
access to the asymptotics.  Moreover, one can check numerically that
our asymptotic formula (\ref{eq:Amean_asympt}) agrees accurately with
the exact expression (\ref{eq:sum_Gauss}) even for moderately large
$n$.

The leading term of Eq. (\ref{eq:Amean_asympt}) was shown to be valid
for a generic random walk with increments of a finite variance (see
Proposition 3.3 in \cite{Wade15}).  Moreover, our numerical
simulations (see Sec. \ref{sec:exp_radial} and
Fig. \ref{fig:exp_radial}) suggest that the obtained formula
(\ref{eq:Amean_asympt}) (including the subleading terms) may be
applicable for some other isotropic processes.  In other words, the
technical assumption about the independence of the jumps along $x$ and
$y$ might be relaxed in future.  This statement, which is uniquely
based on numerical simulations for some jump distributions, is
conjectural.  In turn, the isotropy assumption is important, as
illustrated by numerical simulations.

The large $n$ asymptotic relations (\ref{eq:Lmean_asympt},
\ref{eq:Ln_Levy1}, \ref{eq:Ln_Levy2}, \ref{eq:Amean_asympt}) are the
main results of the paper.  Setting $t = n\tau$ and $D =
2\sigma^2/(4\tau)$ with a time step $\tau$, one recovers from
Eqs. (\ref{eq:Lmean_asympt}, \ref{eq:Amean_asympt}) the same leading
terms as in Eq. (\ref{eq:perim_BM}, \ref{eq:area_BM}) for Brownian
motion (note that we write $2\sigma^2$ in $D$ because $\sigma^2$ is
the variance of jumps along one direction).  It is thus not surprising
that the leading term in Eq. (\ref{eq:Lmean_asympt}) is universal
because its derivation is valid for any planar random walk with a
symmetric and continuous jump distribution and having a finite
variance $\sigma^2$.

Thinking of Brownian motion as a limit of random walks, the subleading
terms in Eq. (\ref{eq:Lmean_asympt}) can be understood as ``finite
size'' corrections.  The first subleading term is valid under the same
assumptions as the leading term, although the coefficient
$\gamma_\theta$ depends on the jump distribution (see examples in
Sec. \ref{sec:simu}).  In turn, the second subleading term depends on
the kurtosis $\K_\theta$ and thus requires an additional assumption
that $\K_\theta$ is finite.

\section{Main steps leading to the derivation of results}
\label{sec:theory}

\subsection{Reduction to a one-dimensional problem}

\begin{figure}
\begin{center}
\includegraphics[width=120mm]{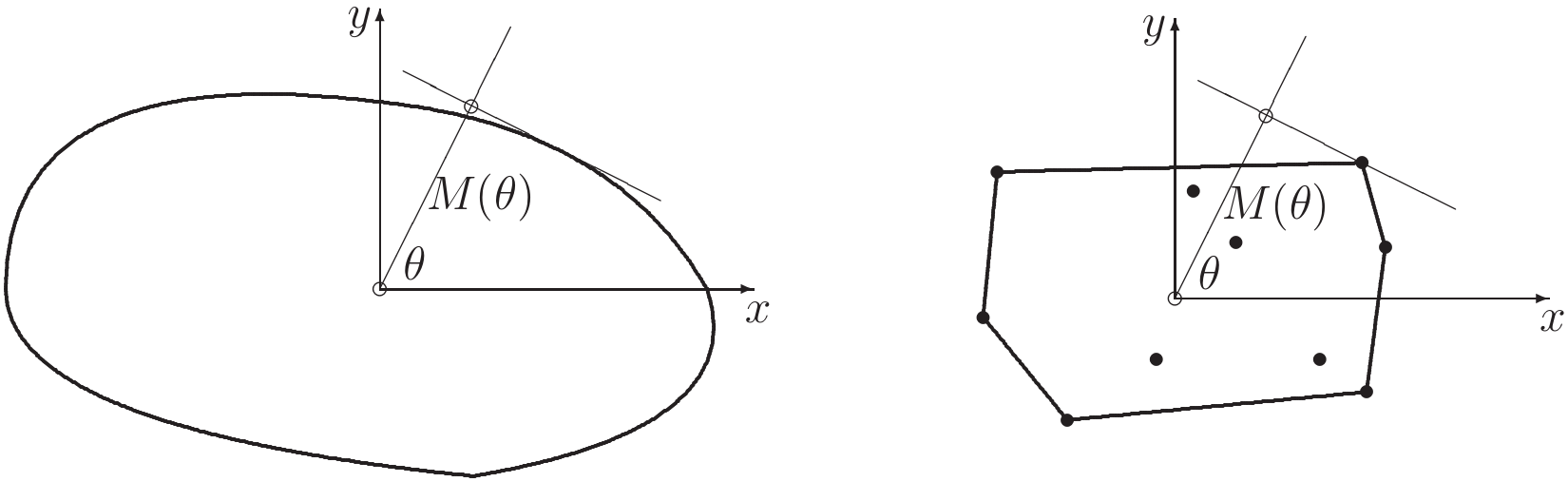}
\end{center}
\caption{
The support function $M(\theta)$ for a closed convex curve (left) and
for a set of points $\{(x_0,y_0),(x_1,y_1),\ldots,(x_n,y_n)\}$
(right).  $M(\theta)$ is the distance between two open circles.}
\label{fig:domain}
\end{figure}

We start with Cauchy's formula for the perimeter $L$ and the area $A$
of an arbitrary convex domain ${\cal C}$ with a reasonably smooth
boundary $\gamma_{\cal C}$~\cite{Cauchy1832,Santalo}.  Let the
boundary $\gamma_{\cal C}$ be parameterized as $(X(s),Y(s))$ with a
curvilinear coordinate $s$ ranging from $0$ to $1$.  Setting the
origin of coordinates inside the domain, one defines the support
function $M(\theta)$ as the distance from the origin to the closest
straight line that does not cross the domain and is perpendicular to
the vector from the origin in direction $\theta$
(Fig. \ref{fig:domain}).  In other words,
\begin{equation}
M(\theta) = \max\limits_{0\leq s \leq 1} \bigl\{ X(s) \cos \theta + Y(s) \sin \theta\bigr\} .
\end{equation}
Cauchy showed that~\cite{Cauchy1832} 
\begin{eqnarray}  
L &=& \int\limits_0^{2\pi } d\theta \, M(\theta) , \\
A &=& \frac12\int\limits_0^{2\pi } d\theta \, \bigl(M^2(\theta) - [M'(\theta)]^2\bigr) .
\end{eqnarray}
For a simple derivation of this formula see Ref.~\cite{Majumdar2010a}.
A straightforward calculation of $M(\theta)$ for a convex hull over a
set of points may seem to be hopeless, as one would need first to
construct the convex hull by identifying and ordering its vertices
among the given set of points and then to compute $M(\theta)$.  The
key idea is that $M(\theta)$ can be found directly from the vertices
of the trajectory as \cite{Spitzer1961,Randon09}
\begin{equation}
M(\theta) = \max\limits_{0\leq k \leq n} \bigl\{ x_k \cos\theta + y_k \sin \theta \bigr\} .
\end{equation}
Moreover, given that the maximum for a fixed $\theta$ is realized by
a certain vertex (with index $k^*$ which discretely changes with
$\theta$), one also obtains the derivative:
\begin{equation}  \label{eq:Mprime}
M'(\theta) = - x_{k^*} \sin\theta + y_{k^*} \cos \theta .
\end{equation}

When the points $(x_k,y_k)$ are random, the perimeter and the area of
the convex hull are random variables.  We focus on the mean values
$\langle L_n\rangle$ and $\langle A_n\rangle$:
\begin{eqnarray}  \label{eq:Lmean}
\langle L_n \rangle & = & \int\limits_0^{2\pi } d\theta \, \langle M(\theta)\rangle , \\
\label{eq:Amean}
\langle A_n \rangle & = & \frac12 \int\limits_0^{2\pi } d\theta \, \bigl(\langle M^2(\theta)\rangle - \langle [M'(\theta)]^2 \rangle\bigr), 
\end{eqnarray}
i.e., the computation is reduced to the first two moments of
$M(\theta)$ and to the mean $\langle [M'(\theta)]^2 \rangle$.  The
important observation is that, for a fixed direction $\theta$, one
needs to characterize the maximum of the projection of points
$(x_k,y_k)$ onto that direction
\begin{equation} 
M(\theta) = \max\limits_{0\leq k \leq n} \{ z_k^\theta\}, \qquad
z_k^\theta = x_k \cos \theta + y_k \sin \theta .
\end{equation}
The projection of a random walk is also a random walk.  In fact, we
can write according to Eq. (\ref{eq:RW_def})
\begin{equation} \label{eq:zk}
z_0^\theta = 0 , \qquad z_k^\theta = z_{k-1}^\theta + \xi_k^\theta  \quad (k=1,2,\ldots,n),
\end{equation}
with
\begin{equation} \label{eq:xik}
\xi_{k}^\theta = \eta_k^x \cos \theta + \eta_k^y \sin \theta. 
\end{equation}
The probability density of $\xi_k^\theta$, $\rho_\theta(z)$, is fully
determined by that of the jump $(\eta_k^x,\eta_k^y)$.  In particular,
its Fourier transform $\hat{\rho}_\theta(k)$ is related to $p(x,y)$ by
Eq. (\ref{eq:hatrho_xi_general}).  The symmetry (\ref{eq:p_symm})
implies that $\hat{\rho}_\theta(-k) = \hat{\rho}_\theta(k)$ and thus
the density $\rho_\theta(z)$ is symmetric.  

Having discussed the general jump distributions, let us mention two
particular cases that will be important later.

\vskip 0.3cm

\noindent (a) in the case of
independent jumps along $x$ and $y$ coordinates, one has $p(x,y) =
p_x(x) p_y(y)$, and thus
\begin{equation}  \label{eq:hatrho_xi}
\hat{\rho}_\theta(k) = \hat{p}_x(k\cos\theta) \, \hat{p}_y(k\sin\theta) ,
\end{equation}
where $\hat{p}_x$ and $\hat{p}_y$ are the Fourier transforms of
$p_x(x)$ and $p_y(y)$, respectively.  

\vskip 0.3cm

\noindent (b) For isotropic jumps,
$p(x,y)$ depends only on the radial coordinate, $p(x,y)dxdy = p_r(r)
dr \, d\phi/(2\pi)$, where $p_r(r)$ is the radial density (that
includes the factor $r$ from the Jacobian).  From
Eq. (\ref{eq:hatrho_xi_general}), one gets
\begin{equation}  \label{eq:hatrho_xi_iso}
\hat{\rho}(k) = \int\limits_0^\infty dr \,  p_r(r) \, J_0(|k|r),  
\end{equation}
in which the integration over the angular coordinate $\phi$ eliminated
the dependence on $\theta$ and resulted in the Bessel function of the
first kind, $J_0(|k|r)$.

\subsection{Formal solution of the one-dimensional problem}

The formal exact solution of the one-dimensional problem can be
obtained via the Pollaczek-Spitzer formula
\cite{Pollaczek52,Spitzer56}.  This formula characterizes the maximum
of partial sums of independent identically distributed random
variables $\xi_k$ with a symmetric and continuous density $\rho(z)$:
\begin{equation}
M_n = \max\{ 0, \xi_1, \xi_1 + \xi_2, \ldots, \xi_1 + \xi_2 + \ldots + \xi_n\} 
\end{equation}
(in this subsection, we temporarily drop the subscript and superscript
$\theta$ from all the variables; it will be restored at the end).
Considering $\xi_k$ as jumps of a random walker, $z_{k} = z_{k-1} +
\xi_k$ (with $z_0 = 0$), one can also write
\begin{equation}
M_n = \max\{ z_0, z_1, z_2, \ldots, z_n\}.  
\end{equation}
Pollaczek and later Spitzer showed that the cumulative distribution
$Q_n(z) = \P\{ M_n \leq z\}$ of $M_n$ satisfies the following identity
for $0 \leq s \leq 1$ and $\lambda
\geq 0$
\begin{equation} \label{eq:PS_identity}
\sum\limits_{n=0}^\infty s^n \, \langle e^{-\lambda M_n} \rangle = 
\sum\limits_{n=0}^\infty s^n \int\limits_0^\infty dz \, e^{-\lambda z} Q'_n(z) = \frac{\phi(s,\lambda)}{\sqrt{1-s}}  \, ,
\end{equation}
with
\begin{equation}  \label{eq:phi}
\phi(s,\lambda) = \exp\left( - \frac{\lambda}{\pi} \int\limits_0^\infty dk \, \frac{\ln(1 - s \hat{\rho}(k))}{\lambda^2 + k^2} \right) ,
\end{equation}
where $Q'_n(z) = dQ_n(z)/dz$ is the probability density of the maximum
\cite{Pollaczek52,Spitzer56}. In principle, all the moments of $M_n$
can be obtained from the formula in Eq. (\ref{eq:PS_identity}).
However, in practice, deriving explicitly the moments of $M_n$ by
inverting this formula is highly nontrivial~\cite{Majumdar09}.  For
example, the expected maximum of a discrete-time random walk $\langle
M_n\rangle$ appears in a number of different problems, from packing
algorithms in computer science~\cite{Coffman98}, all the way to the
survival probability of a single or multiple walkers in presence of a
trap~\cite{Comtet05,MCZ2006,ZMC2007,ZMC2009,Franke2012,MMS2017}.  It
has also appeared in the context of the order, gap and record
statistics of random
walks~\cite{SM2012,SM_review,MMS2013,WMS2012,GMS2017} and $\langle
M_n\rangle$ has been analyzed for large $n$ (for the leading and the
next subleading term) in detail using the Pollaczek-Spitzer formula in
Eq. (\ref{eq:PS_identity}).  Here, in addition to calculating the
first three terms in the asymptotic expansion of $\langle M_n\rangle$
for $n\gg 1$, we also calculate the large $n$ behavior of the second
moment $\langle M_n^2 \rangle$, that we need for the computation of
the mean area of the convex hull.

In fact, the Pollaczek-Spitzer formula also determines the generating
functions for all moments of $M_n$:
\begin{equation}  \label{eq:hm}
h_m(s) = \sum\limits_{n=0}^\infty s^n \, \langle M_n^m \rangle =
(-1)^m \lim\limits_{\lambda\to 0} \frac{\partial^m}{\partial \lambda^m} \frac{\phi(s,\lambda)}{\sqrt{1-s}}  \, .
\end{equation}
By considering a general asymptotic expansion
\begin{equation}
\hat{\rho}(k) \simeq 1 - |ak|^\mu + o(|k|^\mu)  \qquad (k\to 0)
\end{equation}
with an exponent $0 < \mu \leq 2$ and a scale $a > 0$, we derive in
\ref{sec:Agenerating} the exact expressions
\begin{equation}  \label{eq:h1}
h_1(s) = \frac{1}{\pi(1-s)} \int\limits_0^\infty \frac{dk}{k^2} \ln \left(\frac{1- s \hat{\rho}(k)}{1-s}\right) \qquad (0 \leq s < 1),
\end{equation}
which is valid for any $1<\mu \le 2$ (note that $\langle M_n\rangle =
\infty$ for $0 < \mu \leq 1$), and
\begin{equation}  \label{eq:h2}
h_2(s) = (1-s) [h_1(s)]^2 + \frac{a^2 s}{(1-s)^2}  \qquad (0 \leq s < 1),
\end{equation}
which is valid for $\mu = 2$ (note that $\langle M_n^2\rangle =
\infty$ for $0 < \mu < 2$).  The exact relations (\ref{eq:h1},
\ref{eq:h2}) are new results which allow one to study the first two
moments of the maximum $M_n$.  In the next subsection, we will analyze
the expansion of Eqs. (\ref{eq:h1}, \ref{eq:h2}) as $s\to 1$ in order
to determine the asymptotic behavior of the moments $\langle
M_n\rangle$ and $\langle M_n^2\rangle$ as $n\to \infty$.  We consider
separately jumps with a finite variance, and L\'evy flights.

\subsection{Mean perimeter and mean area of the convex hull}

\subsubsection{Mean perimeter for jumps with a finite variance.}

Given a generic continuous jump distribution $p(x,y)$ satisfying the
property in Eq. (\ref{eq:p_symm}), we determine $\hat{\rho}_\theta(k)$
using Eq. (\ref{eq:hatrho_xi_general}).  Furthermore, by examining the
small $k$ behavior of $\hat{\rho}_\theta(k)$, we determine the
$\theta$-dependent variance $\sigma_\theta^2$ and the
$\theta$-dependent kurtosis $\K_\theta$, using respectively Eqs.
(\ref{eq:sigma_theta}) and (\ref{eq:a4}).  In addition, knowing
$\hat{\rho}_\theta(k)$ from Eq.  (\ref{eq:hatrho_xi_general}), we also
determine $\gamma_\theta$ in Eq. (\ref{eq:gamma}).  Equipped with
these three quantities $\sigma_\theta$, $\K_\theta$ and
$\gamma_\theta$, we show in \ref{sec:Aderivation} that the leading
large $n$ terms of the first two moments of $M_n$ are given by
\begin{eqnarray}  \label{eq:Mn}
\frac{\langle M_n \rangle}{\sigma_\theta} &\simeq& 
\frac{\sqrt{2}}{\sqrt{\pi}}\, n^{\frac12} + \gamma_\theta 
+ \frac{\K_\theta + 3}{12\sqrt{2\pi}}\, n^{-\frac12} + o(n^{-\frac12}), \\
\label{eq:Mn2}
\frac{\langle M_n^2\rangle}{\sigma_\theta^2} &\simeq& n + 
\frac{\sqrt{8}\, \gamma_\theta}{\sqrt{\pi}}\, n^{\frac 12} + (\K_\theta/12 + 
\gamma_\theta^2) + o(1) \, .
\end{eqnarray}
For the mean perimeter of the convex hull, we will only need the first
moment in Eq. (\ref{eq:Mn}).  Indeed, using Eq. (\ref{eq:Lmean}), the
integration of the expansion (\ref{eq:Mn}) over $\theta$ from $0$ to
$2\pi$ yields the announced result (\ref{eq:Lmean_asympt}) for the
mean perimeter of the convex hull.  The result for the second moment
in Eq. (\ref{eq:Mn2}) will be needed later to determine the mean area
$\langle A_n\rangle$.

\subsubsection{Mean perimeter for L\'evy flights.}
\label{sec:Levy}

When the variance of jumps is infinite, one gets the Taylor expansion
Eq. (\ref{eq:hatrho_mu}), with the scaling exponent $0 < \mu < 2$.
When $0 < \mu \leq 1$, the mean perimeter of the convex hull is
infinite.  Throughout this section, we focus on the case $1 < \mu <
2$, in which the first moment of jumps is finite (and zero due to the
assumption of a symmetric distribution), whereas the variance is
infinite.  In this case, the leading behavior of the mean maximum of
partial sums is universal \cite{Comtet05}
\begin{equation}  \label{eq:Mn_mu}
\langle M_n \rangle \simeq a_\theta \frac{ \mu \Gamma(1- 1/\mu)}{\pi} \, n^{1/\mu} + o(n^{1/\mu}) \qquad (n\gg 1).
\end{equation}
However, the subleading term depends on finer details of the jump
distribution.  In order to determine the subleading term, we consider
the expansion (\ref{eq:hatrho_nu}) with the subleading term $b_\theta
|k|^\nu$ such that $\nu > \mu$.  We distinguish two cases: $\mu < \nu
< \mu+1$ and $\nu > \mu + 1$.  In \ref{sec:ALevy}, we derive the
following asymptotics results:
\begin{equation}  \label{eq:Mn_Levy1}
\fl
\langle M_n\rangle \simeq a_\theta \, \frac{\mu \Gamma(1- 1/\mu)}{\pi} \, n^{1/\mu}  
- a_\theta^{1-\nu} \, b_\theta \, \frac{\Gamma((\nu-1)/\mu)}{\pi (\mu+1-\nu)} \, n^{1-(\nu-1)/\mu} + o(n^{1-(\nu-1)/\mu})  \quad (n\gg 1)
\end{equation}
for $\mu < \nu < \mu+1$, and 
\begin{equation}  \label{eq:Mn_Levy2}
\langle M_n\rangle \simeq a_\theta \, \frac{\mu \Gamma(1- 1/\mu)}{\pi} \, n^{1/\mu} + \gamma_\theta + o(1)  \quad (n\gg 1)
\end{equation}
for $\nu > \mu + 1$, with $\gamma_\theta$ given by
Eq. (\ref{eq:gamma_Levy2}).  The asymptotic relation
(\ref{eq:Mn_Levy2}) was first derived in \cite{Comtet05} for the
particular case $\nu = 2\mu$.  One can see that for $\mu < \nu <
\mu+1$, the subleading term of $\langle M_n\rangle$ grows with $n$,
whereas for when $\nu > \mu + 1$, the subleading term is constant.
Higher-order corrections can be derived as well.

Finally, using the Cauchy formula (\ref{eq:Lmean}), the integration of
Eqs. (\ref{eq:Mn_Levy1}, \ref{eq:Mn_Levy2}) over $\theta$ from $0$ to
$2\pi$ yields Eqs. (\ref{eq:Ln_Levy1}, \ref{eq:Ln_Levy2}) for the mean
perimeter of the convex hull, announced in Section \ref{sec:outline}.

\subsubsection{Mean area for isotropic Gaussian jumps.}

According to Eq. (\ref{eq:Amean}), the expansion (\ref{eq:Mn2})
determines the first contribution to the mean area.  This contribution
was calculated for an arbitrary symmetric continuous jump distribution
with a finite variance.  The second contribution to the mean area
comes from $\langle [M'(\theta)]^2\rangle$ that has to be computed
separately.  We recall that $M'(\theta)$ is given by
Eq. (\ref{eq:Mprime}).  Our computation of this contribution relies on
two additional simplifying assumptions: (a) the jumps along $x$ and
$y$ coordinates are independent and (b) the jump process is isotropic,
i.e., the distribution of jumps does not depend on their direction.
In this case, using the isotropy condition (b), we get
\begin{eqnarray}
\langle [M(\theta)]^2\rangle &=& \langle [M(0)]^2\rangle = \langle x_{k^*}^2 \rangle , \\
\langle [M'(\theta)]^2\rangle &=& \langle [M'(0)]^2\rangle = \langle y_{k^*}^2 \rangle  .
\end{eqnarray}
The disentanglement of $\langle [M(\theta)]^2\rangle$ and $\langle
[M'(\theta)]^2\rangle$ allows one to compute the latter one by using
the following argument.  We recall that $k^*$ is the index of the
maximal position among $x_k$, i.e., its statistics is fully determined
by the jumps along $x$ coordinate.  Once this statistics is known, the
mean $\langle [M'(0)]^2\rangle$ can be found by taking the conditional
expectation of $y^2_{k^*}$ at any fixed value $k^*$ and then the
expectation with respect to the distribution of $k^*$. Now, once we
condition $k^*$, i.e., the time step at which the $x_k$'s achieve
their maximum, the $y_k$ process will, in general, be affected by this
conditioning. However, if $x_k$ and $y_k$ are independent (which
happens when the jump process satisfies property (a) above), we get
\begin{equation}
\langle [M'(0)]^2\rangle = \sigma^2 \, \langle k^* \rangle ,
\end{equation}
where $\langle [\eta_k^y]^2\rangle = \sigma^2$.  It remains to find
$\langle k^* \rangle$.  For symmetric and continuous jump
distributions, it follows from symmetry that $\langle k^* \rangle=n/2$
independent of the details of the jump PDF.  This can be deduced
formally also, by noting that the time step $k^*$, at which the
maximum of $x_k$ is achieved, has a universal distribution,
independent of the jump distribution (given that the latter is
continuous and symmetric)~\cite{Majumdar09}:
\begin{equation}
P_n(k^* = k) = \left(\begin{array}{c} 2k \\ k \end{array}\right)  \left(\begin{array}{c} 2(n-k) \\ n-k \end{array}\right) 2^{-2n} .
\end{equation}
This is the direct consequence of the Sparre Andersen theorem.  From
this distribution, one easily computes the mean value as
\begin{equation}
\langle k^* \rangle = \frac{n}{2} \, ,
\end{equation}
and thus
\begin{equation}  \label{eq:Mprime_average}
\langle [M'(\theta)]^2\rangle = \sigma^2 \, \frac{n}{2} \,.
\end{equation}
This yields Eq. (\ref{eq:Amean_asympt}) for the mean area of the
convex hull for an isotropic jump process with independent jumps along
$x$ and $y$ coordinates.  As discussed in Sec. \ref{sec:outline}, the
only process satisfying two requirements (a) and (b) is the isotropic
Gaussian process.  Our formula (\ref{eq:Amean_asympt}) is thus
provided only in this case.  For a more general case, one would need
to compute the joint distribution of the maximum position $k^*$ and
both values $x_{k^*}$ and $y_{k^*}$ which is more complicated and
remains an open problem.

\section{Examples and simulations}
\label{sec:simu}

In this section, we illustrate the above general results on several
examples of symmetric planar random walks.  We derive the explicit
values of the relevant parameters that determine the mean perimeter
and the mean area.  We also investigate the effect of anisotropy of
the jump distribution on the convex hull properties.  Finally, we
compare our theoretical predictions to the results of Monte Carlo
simulations.  For this purpose, we generate $10^5$ planar random
walks, compute the convex hull for each generated trajectory with
$n+1$ points by using the Matlab function 'convhull', and determine
its perimeter and area.  These simulations yield the representative
statistics of perimeters and areas from which the mean values are
computed.

\subsection{Gaussian jumps}

We first consider the basic example of Gaussian jumps which are
independent along $x$ and $y$ coordinates and characterized by
variances $\sigma_x^2$ and $\sigma_y^2$.  Substituting the jump
probability density,
\begin{equation}
p(x,y) = \frac{\exp\bigl(-\frac{x^2}{2\sigma^2_x}\bigr)}{\sqrt{2\pi}\, \sigma_x} \, 
\frac{\exp\bigl(-\frac{y^2}{2\sigma^2_y}\bigr)}{\sqrt{2\pi} \, \sigma_y} \, ,
\end{equation}
into Eq. (\ref{eq:hatrho_xi_general}) yields $\hat{\rho}_\theta(k) =
e^{-k^2 \sigma^2_\theta/2}$, with $\sigma_\theta^2 = \sigma_x^2
\cos^2\theta + \sigma_y^2 \sin^2\theta$.  One can see that the
anisotropy only affects the variance $\sigma_\theta^2$, whereas the
two other relevant parameters, $\gamma$ and $\K$, which are rescaled
by variance, do not depend on $\theta$.  One finds $\K = 3$, whereas
the integral in Eq. (\ref{eq:gamma}) was computed exactly in
\cite{Comtet05} and provided in Eq. (\ref{eq:gamma_Gauss}).
Assuming (without loss of generality) that $\sigma_x \geq \sigma_y$,
we set
\begin{equation}  \label{eq:sigma_Gauss}
\sigma \equiv \frac{1}{2\pi} \int\limits_0^{2\pi} d\theta \, \sigma_\theta = \frac{2}{\pi} E\biggl(\sqrt{1 - (\sigma_y/\sigma_x)^2}\biggr) \, \sigma_x ,
\end{equation}
where $E(k)$ is the complete elliptic integral of the second kind (for
the isotropic case, $\sigma_x = \sigma_y = \sigma$).  With this
notation, we get the expansion coefficients
\begin{equation}  \label{eq:Cj_Gauss}
\sigma^{-1} C_0 = \sqrt{8\pi} , \qquad  \sigma^{-1} C_1 = 2\pi \gamma = -3.6605 \ldots , \qquad
\sigma^{-1} C_2 = \frac{\sqrt{\pi}}{\sqrt{2}} \,.
\end{equation}

\begin{figure}
\begin{center}
\includegraphics[width=75mm]{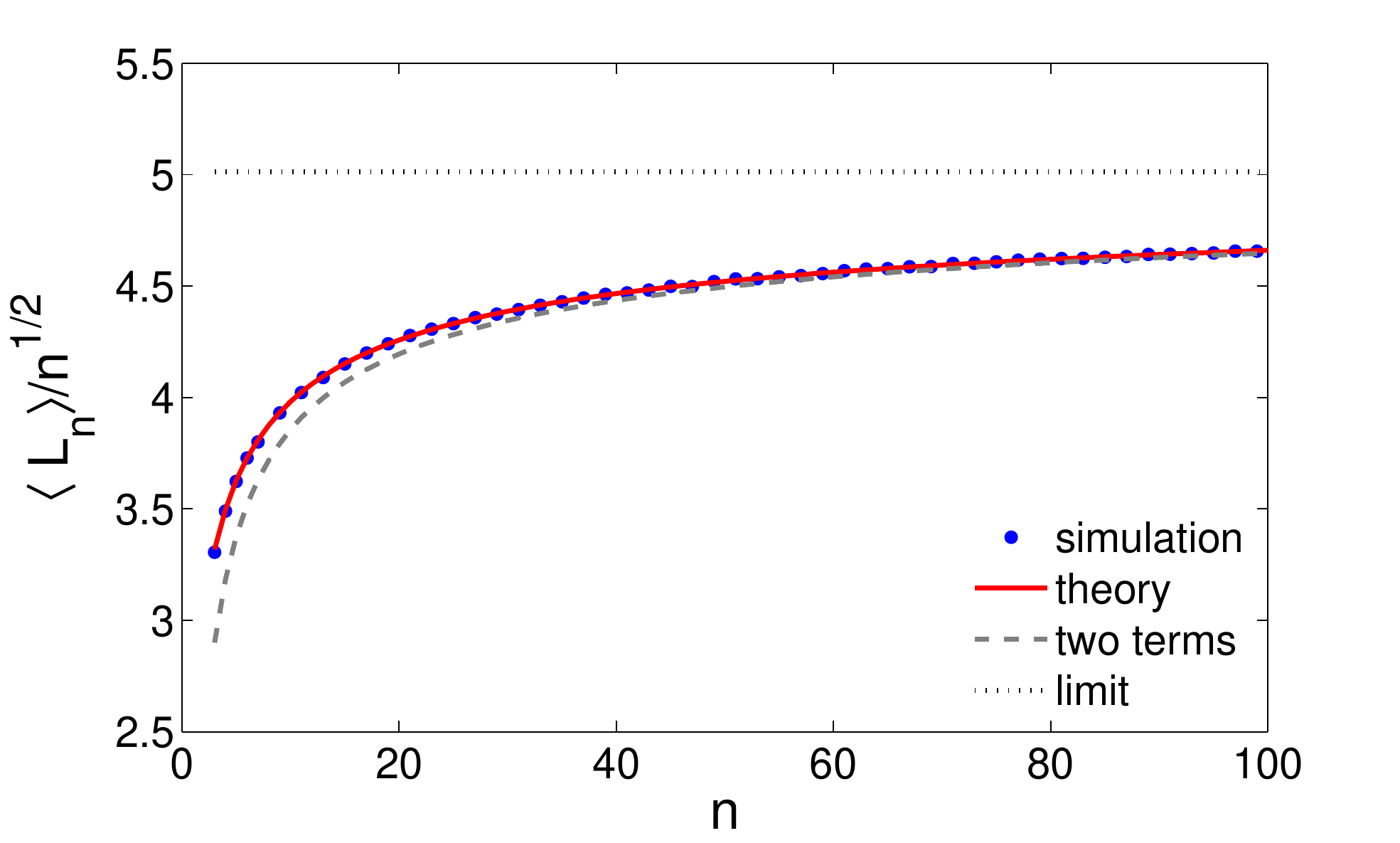}  
\includegraphics[width=75mm]{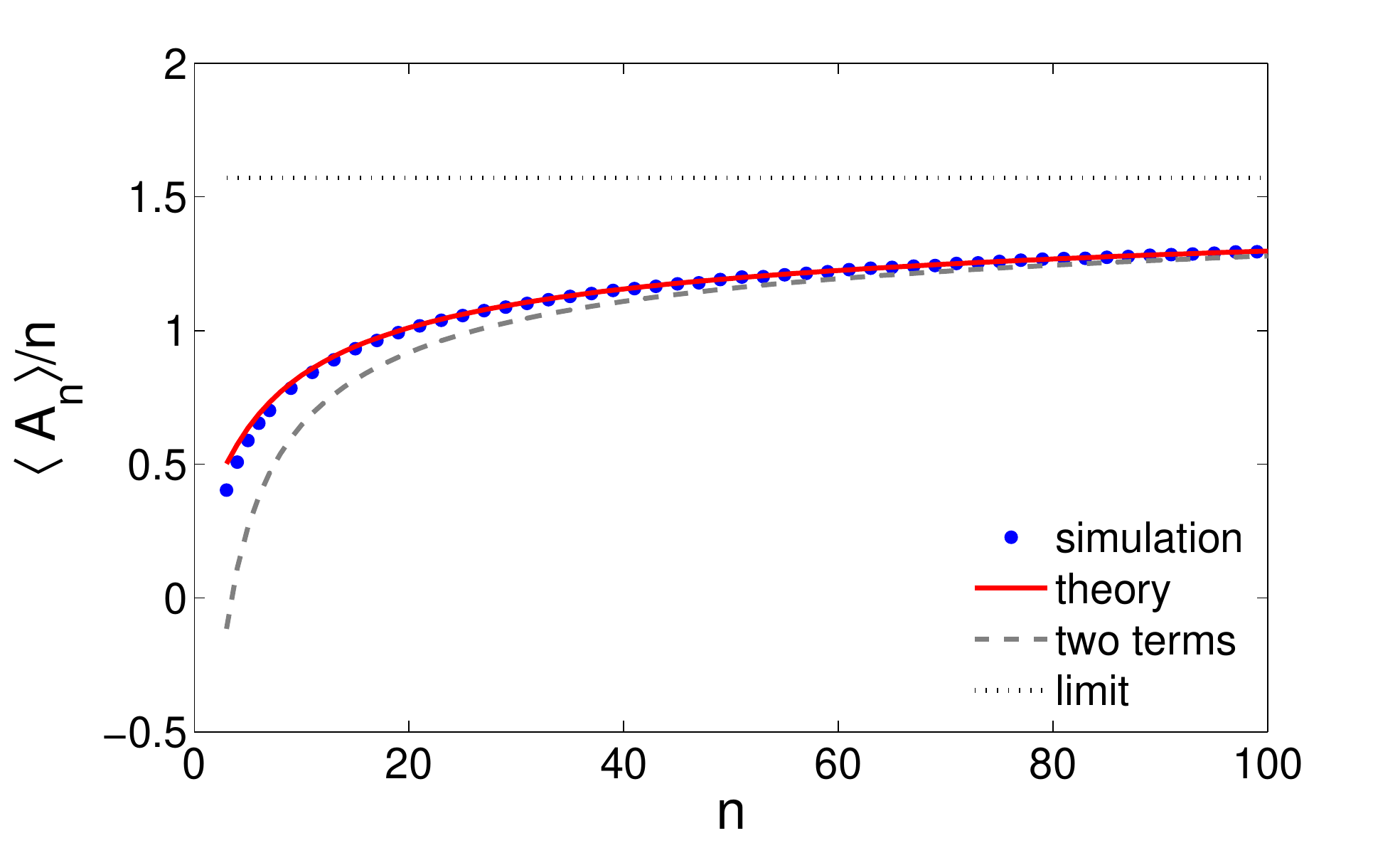}  
\end{center}
\caption{
The rescaled mean perimeter, $\langle L_n\rangle/n^{1/2}$, (left), and
the rescaled mean area, $\langle A_n\rangle/n$, (right), for isotropic
planar random walks with independent Gaussian jumps, with $\sigma_x =
\sigma_y = 1$.  The results of Monte Carlo simulations (shown by
circles) are in perfect agreement with our theoretical predictions
(\ref{eq:Lmean_asympt}, \ref{eq:Amean_asympt}) (shown by solid line).
The dotted horizontal line presents the coefficients $\sqrt{8\pi}$ and
$\pi/2$ of the leading term, whereas the dashed line illustrates the
theoretical predictions with only two principal terms. }
\label{fig:Gauss_iso}
\end{figure}

Figure \ref{fig:Gauss_iso} shows the rescaled mean perimeter, $\langle
L_n\rangle/n^{1/2}$, and the rescaled mean area, $\langle
A_n\rangle/n$, for isotropic planar random walk with independent
Gaussian jumps ($\sigma_x = \sigma_y = 1$).  The results of Monte
Carlo simulations are in perfect agreement with our theoretical
predictions (\ref{eq:Lmean_asympt}, \ref{eq:Amean_asympt}).  One can
see that the subleading terms play an important role.  In fact, if one
kept only the leading term and omitted the subleading terms, one would
get the horizontal dotted line.  This corresponds to the case of
Brownian motion, in which only the leading term is present, see
Eqs. (\ref{eq:perim_BM}, \ref{eq:area_BM}).  The subleading terms
account for the discrete-time character of random walks which is
particularly important for moderate values of $n$.  In order to
highlight the role of the third term in the asymptotic expansions, we
also draw by dashed line the asymptotics without this term.  As
expected, the third term improves the quality of the theoretical
prediction at small $n$.  Note also that the asymptotic relation is
slightly less accurate for the mean area.

\begin{figure}
\begin{center}
\includegraphics[width=75mm]{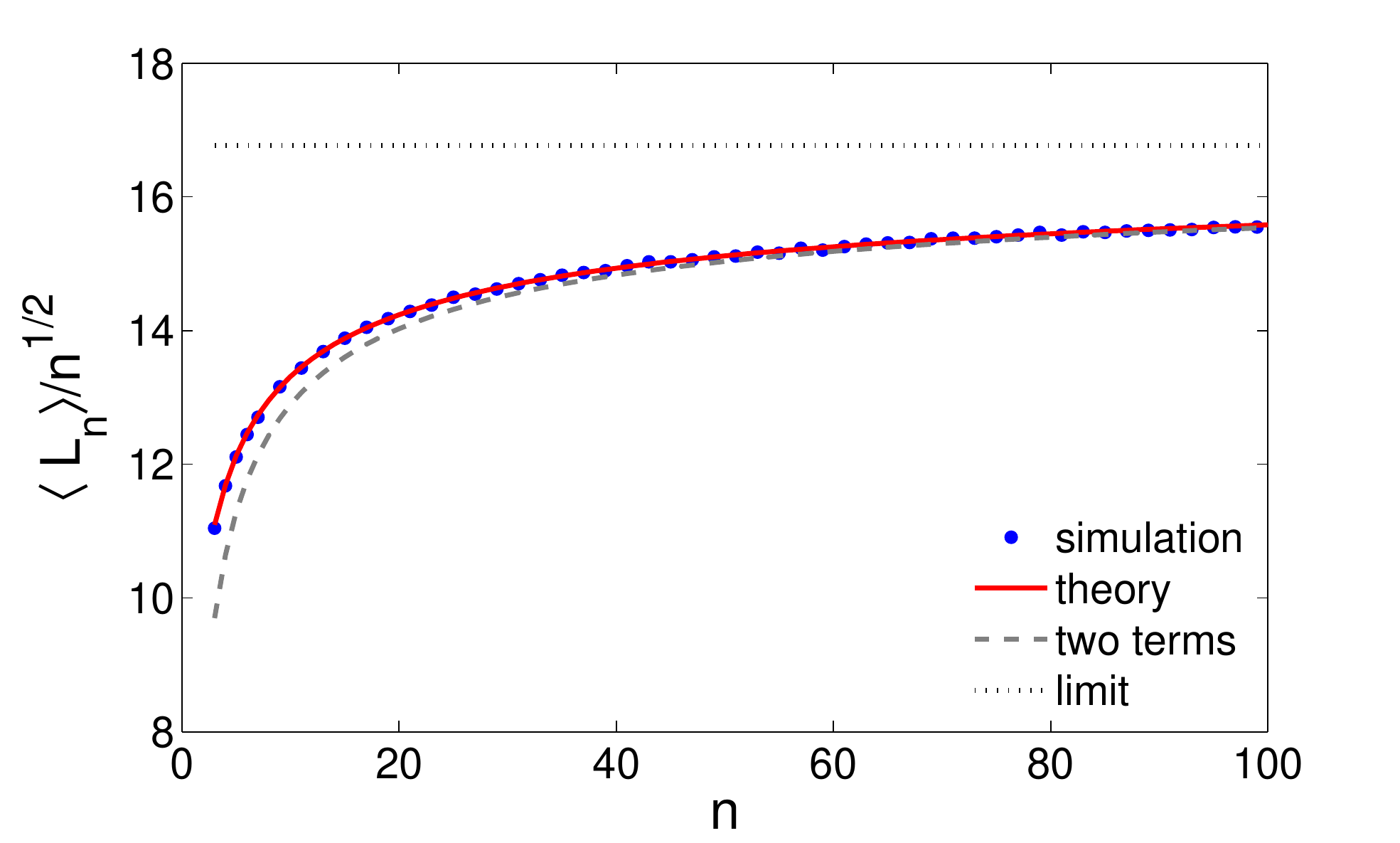}  
\end{center}
\caption{
The rescaled mean perimeter, $\langle L_n\rangle/n^{1/2}$, for
anisotropic planar random walk with independent Gaussian jumps, with
$\sigma_x = 5$ and $\sigma_y = 1$.  The results of Monte Carlo
simulations (shown by circles) are in perfect agreement with our
theoretical prediction (\ref{eq:Lmean_asympt}) (shown by solid line).
The dotted horizontal line presents the coefficient $\sigma
\sqrt{8\pi}$ of the leading term (with $\sigma = 5 \frac{2}{\pi}
E\bigl(\sqrt{1-1/25}\bigr) = 3.3439\ldots$), whereas the dashed line
illustrates the theoretical predictions with only two principal
terms.}
\label{fig:Gauss_ani}
\end{figure}

In Fig. \ref{fig:Gauss_ani}, we consider the convex hull for
anisotropic random walk with independent Gaussian jumps, with
$\sigma_x = 5$ and $\sigma_y = 1$.  In this case, one can use the
asymptotic formula (\ref{eq:Lmean_asympt}), in which the expansion
coefficients $C_j$ are given by Eq. (\ref{eq:Cj_Gauss}), with an
effective variance $\sigma^2$ computed in Eq.  (\ref{eq:sigma_Gauss}).
In this example, $\sigma = 5 \frac{2}{\pi} E\bigl(\sqrt{1-1/25}\bigr)
= 3.3439\ldots$.  For the mean perimeter, one observes a perfect
agreement between the theoretical predictions and Monte Carlo
simulations.  In turn, our asymptotic formula (\ref{eq:Amean_asympt})
for the mean area is not applicable for anisotropic case, as also
confirmed by simulations (not shown).

\subsection{Exponentially distributed radial jumps}
\label{sec:exp_radial}

The next common model has exponentially distributed radial jumps with
uniform angular distribution.  This is a particular realization of a
``run-and-tumble'' model of bacterial motion
\cite{Berg72,Berg,Lauga09}.  Substituting the radial density $p_r(r) =
\sigma^{-1}\, e^{-r/\sigma}$ into Eq. (\ref{eq:hatrho_xi_iso}) yields
$\hat{\rho}(k) = \bigl(1 + (k\sigma)^2\bigr)^{-1/2}$.  One gets $\K =
9$ and
\begin{equation} 
\gamma = \frac{1}{\pi \sqrt{2}} \int\limits_0^\infty \frac{dk}{k^2} \ln \biggl(\frac{1 - \bigl(1 + 2k^2\bigr)^{-1/2}}{k^2}\biggr) 
= -0.8183\ldots 
\end{equation}
from which the expansion coefficients are
\begin{equation}
\sigma^{-1} C_0 = \sqrt{8\pi} , \qquad \sigma^{-1} C_1 = - 5.1416\ldots ,  \qquad \sigma^{-1} C_2 = \sqrt{2\pi} .
\end{equation}

\begin{figure}
\begin{center}
\includegraphics[width=75mm]{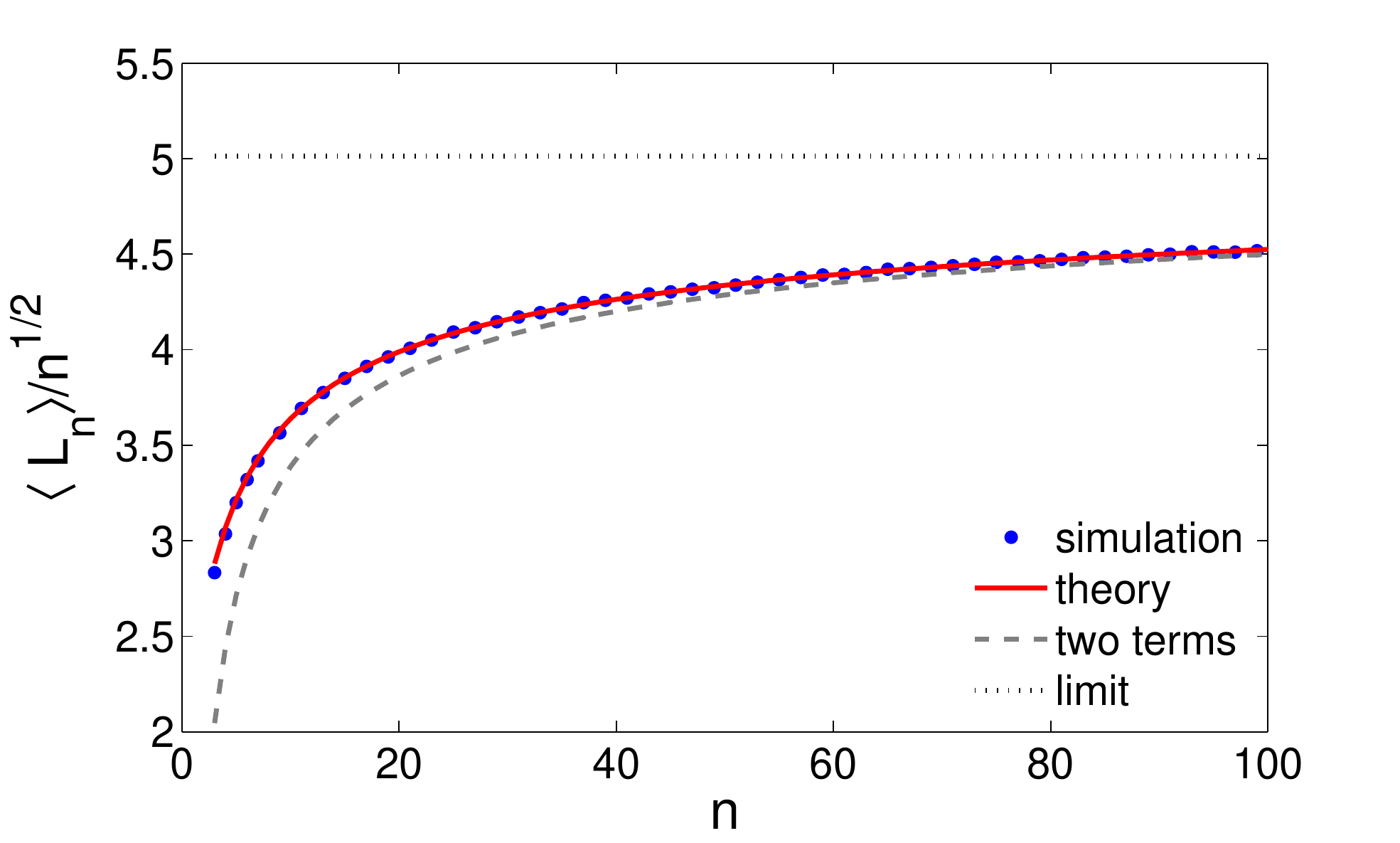} 
\includegraphics[width=75mm]{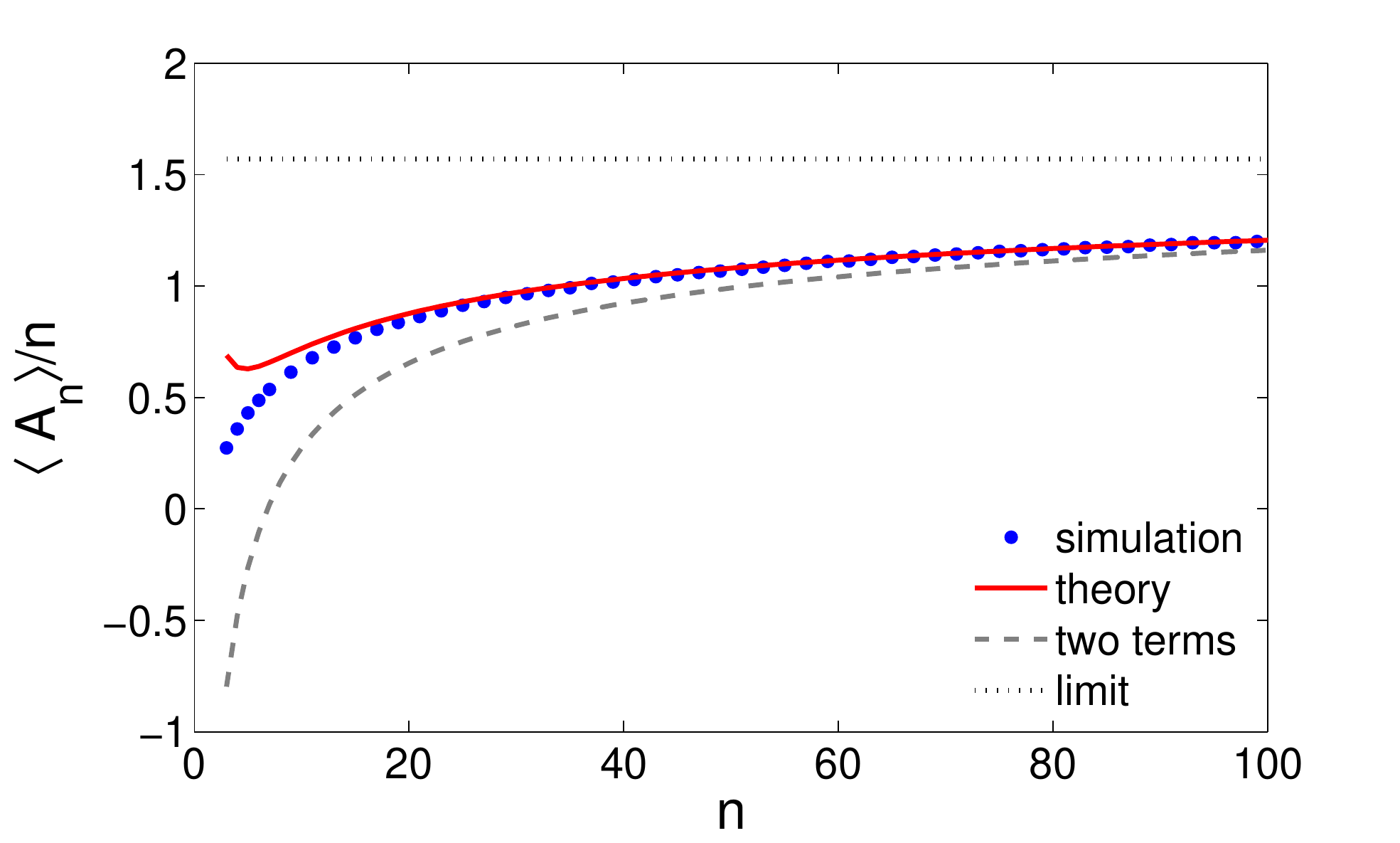} 
\end{center}
\caption{
The rescaled mean perimeter, $\langle L_n\rangle/n^{1/2}$, (left), and
the rescaled mean area, $\langle A_n\rangle/n$, (right), for isotropic
planar random walk with exponentially distributed radial jumps, with
$\sigma = 1$.  The results of Monte Carlo simulations (shown by
circles) are in perfect agreement with our theoretical predictions
(\ref{eq:Lmean_asympt}, \ref{eq:Amean_asympt}) (shown by solid line).
The dotted horizontal line presents the coefficient $\sqrt{8\pi}$ and
$\pi/2$ of the leading term, whereas the dashed line illustrates the
theoretical predictions with only two principal terms.}
\label{fig:exp_radial}
\end{figure}

Figure \ref{fig:exp_radial} illustrates the obtained results.  As for
isotropic Gaussian jumps, there is a perfect agreement between theory
and simulations for the mean perimeter.  We also present the results
for the mean area.  We recall that the theoretical formula
(\ref{eq:Amean_asympt}) was derived under the assumption of
independent jumps along $x$ and $y$ coordinates, which evidently fails
for the exponential radial jumps.  In spite of this failure, the
theoretical formula (\ref{eq:Amean_asympt}) is in perfect agreement
with Monte Carlo simulations, except for very small $n$.  This
empirical observation suggests a possibility to relax this technical
assumption in future, at least for large $n$.

\subsection{Independent exponentially distributed jumps}

We also consider the case, when the jumps along $x$ and $y$
coordinates are independent and exponentially distributed, with two
densities $p_x(x) = \frac12 \sigma_x^{-1} e^{-|x|/\sigma_x}$ and
$p_y(y) = \frac12 \sigma_y^{-1} e^{-|y|/\sigma_y}$.  The Fourier
transforms are $\hat{p}_x(k) = (1 + (k\sigma_x)^2)^{-1}$ and
$\hat{p}_y(k) = (1 + (k\sigma_y)^2)^{-1}$ so that
\begin{equation}
\hat{\rho}_\theta(k) = \frac{1}{1 + (k\sigma_x)^2 \cos^2 \theta} \, \frac{1}{1 + (k\sigma_y)^2 \sin^2 \theta} \,. 
\end{equation}
We get $\sigma^2_\theta = 2(\sigma_x^2 \cos^2\theta + \sigma_y^2
\sin^2\theta)$ and
\begin{equation}
\K_\theta = 24 \frac{\sigma_x^4 \cos^4\theta + \sigma_x^2 \sigma_y^2 \sin^2\theta \cos^2\theta + \sigma_y^4 \sin^4\theta}{\sigma_\theta^4} \,.
\end{equation}
Using the identity (\ref{eq:auxil1}), we compute explicitly
\begin{equation}
\gamma_\theta = \frac{\sqrt{2} \sigma_x \sigma_y |\sin\theta \cos\theta| - \sigma_\theta (\sigma_x |\cos\theta| + \sigma_y |\sin\theta|)}{\sigma_\theta^2} .
\end{equation}

\begin{figure}
\begin{center}
\includegraphics[width=75mm]{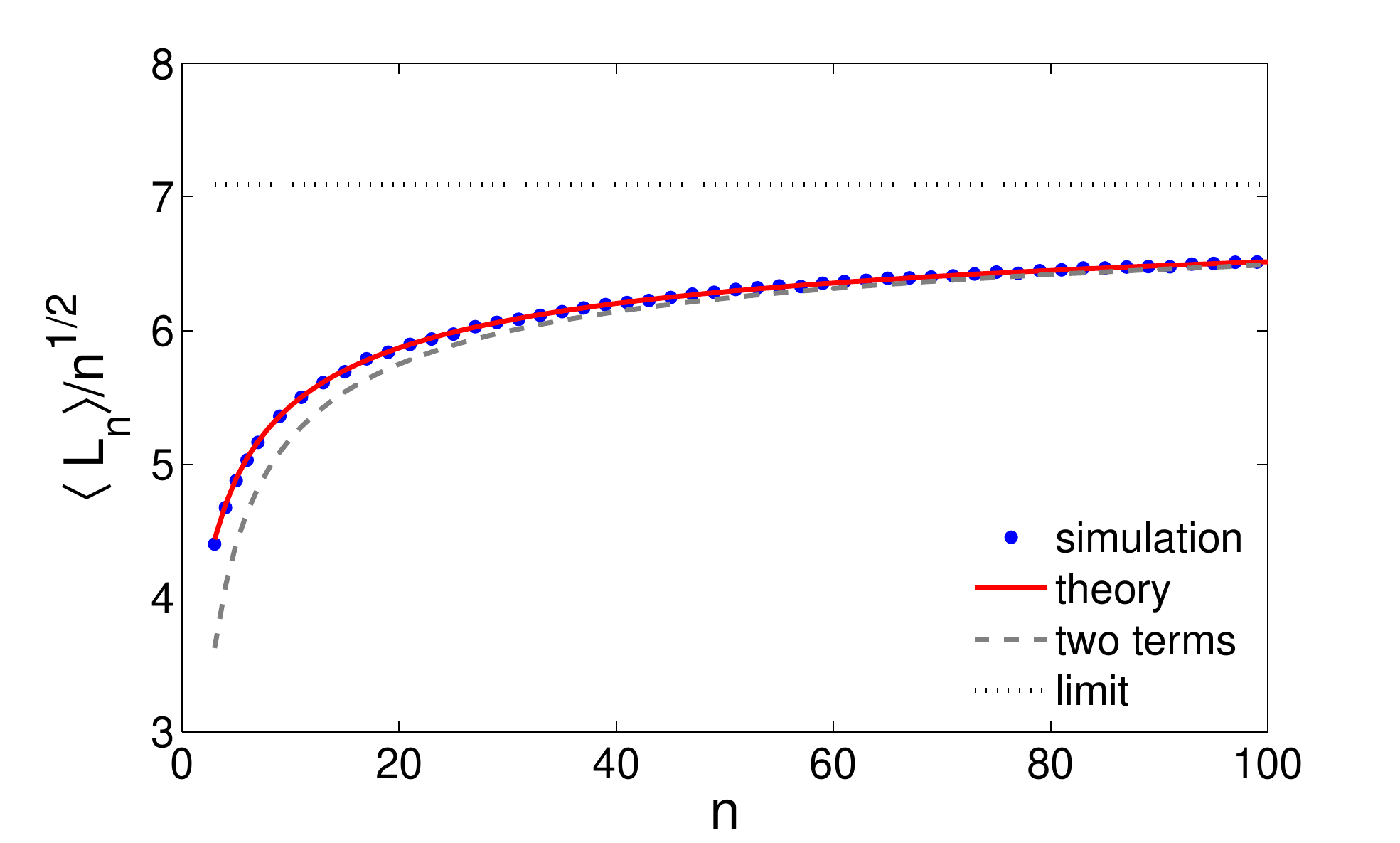}  
\end{center}
\caption{
The rescaled mean perimeter, $\langle L_n\rangle/n^{1/2}$, for
anisotropic planar random walk with independent exponentially
distributed jumps, with $\sigma_x = \sigma_y = 1$.  The results of
Monte Carlo simulations (shown by circles) are in perfect agreement
with our theoretical prediction (\ref{eq:Lmean_asympt}) (shown by
solid line).  The dotted horizontal line presents the coefficient
$4\sqrt{\pi}$ of the leading term, whereas the dashed line illustrates
the theoretical predictions with only two principal terms.}
\label{fig:exp_indep}
\end{figure}

In the particular case $\sigma_x = \sigma_y = \sigma$, one gets
$\sigma^2_\theta = 2\sigma^2$, $\K_\theta = 6(1 - \sin^2\theta
\cos^2\theta)$, and
\begin{equation}
\gamma_\theta = \frac{|\sin\theta \cos\theta| - |\cos\theta| - |\sin\theta|}{\sqrt{2}} \,.
\end{equation}
For this case, we obtain from Eqs. (\ref{eq:CLn})
\begin{equation}
\sigma^{-1} C_0 = 4\sqrt{\pi} , \qquad     \sigma^{-1} C_1 = -6 , \qquad    \sigma^{-1} C_2 = \frac{11\sqrt{\pi}}{8}\,.
\end{equation}
Figure \ref{fig:exp_indep} illustrates the excellent agreement between
theory and simulations.

\subsection{Radial L\'evy jumps}

\begin{figure}
\begin{center}
\includegraphics[width=85mm]{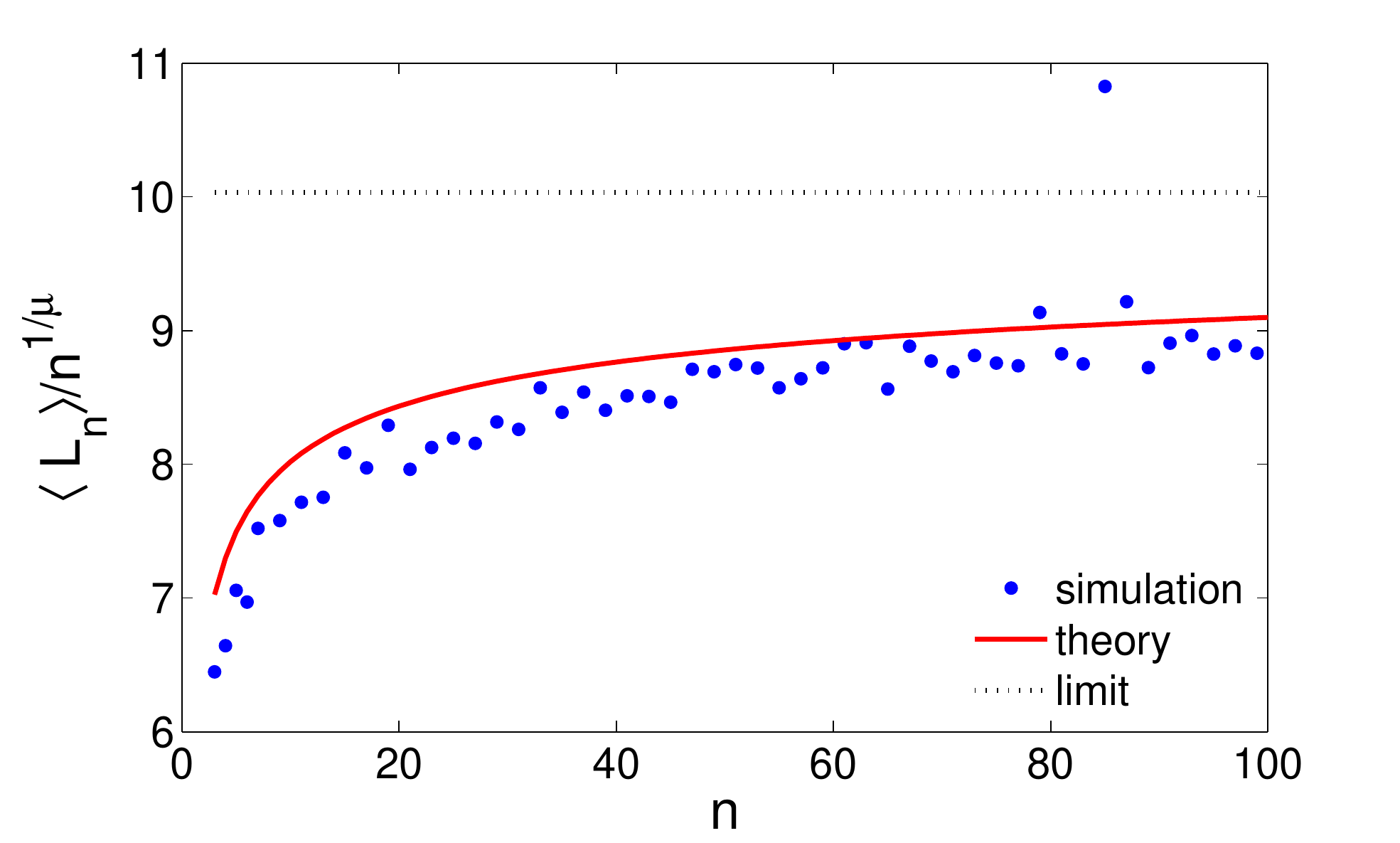}  
\end{center}
\caption{
The rescaled mean perimeter, $\langle L_n\rangle/n^{1/\mu}$, for
isotropic planar random walk with radial L\'evy flights whose lengths
are distributed according to Eq. (\ref{eq:Levy_distrib}) with $\mu =
1.5$ and $R = 1$.  The results of Monte Carlo simulations (shown by
circles) agree well with the theoretical prediction
(\ref{eq:Ln_Levy1}) (shown by solid line). }
\label{fig:levy_radial}
\end{figure}

Now, we investigate the example of radial L\'evy flights with infinite
variance (but finite mean) and uniform angle distribution.  Among
various heavy-tailed jump distributions (e.g., Pareto distributions),
we choose for our illustrative purposes the distribution
\begin{equation}  \label{eq:Levy_distrib}
\P\{ \xi > r\} = \bigl(1 + (r/R)^2\bigr)^{-\alpha}  ,
\end{equation}
with a scale $R > 0$ and the scaling exponent $\mu = 2\alpha$, with
$\frac12 < \alpha < 1$.  For this distribution,
Eq. (\ref{eq:hatrho_xi_iso}) yields a simple closed formula
\cite{Gradshteyn}
\begin{equation}
\hat{\rho}(k) = \frac{2^{1-\alpha}}{\Gamma(\alpha)} (|k|R)^{\alpha} K_{\alpha}(|k|R) ,
\end{equation}   
where $K_\alpha(z)$ is the modified Bessel function of the second kind.
The asymptotic behavior of $K_\alpha(z)$ near $z$ implies as $k\to 0$
\begin{equation}
\hat{\rho}(k) \simeq 1 - \frac{\pi \, (|k|R)^{2\alpha}}{2^{2\alpha} \sin(\pi\alpha) \Gamma(\alpha)\Gamma(\alpha+1)} 
+ \frac{(kR)^2}{4(1-\alpha)} + O(|k|^{2+2\alpha}) .
\end{equation}
Comparing this expansion to Eq. (\ref{eq:hatrho_nu}), one can identify
\begin{equation}  \label{eq:aR}
\fl
a = R \left(\frac{\pi}{2^{2\alpha} \sin(\pi\alpha) \Gamma(\alpha)\Gamma(\alpha+1)}\right)^{\frac{1}{2\alpha}} \,, \qquad
b = \frac{R^2}{4(1-\alpha)} \,, \qquad \nu = 2.
\end{equation}
In Fig. \ref{fig:levy_radial}, the mean perimeter computed by Monte
Carlo simulations for $\mu = 1.5$ and $R = 1$ is compared to the
theoretical prediction (\ref{eq:Ln_Levy1}).  One can see that the
agreement is good but worse than for the earlier examples with a
finite variance.  One obvious reason is that here we have determined
only two terms, whereas Eq. (\ref{eq:Lmean_asympt}) contains three
terms.

\subsection{Independent L\'evy $\alpha$-stable symmetric jumps}

Finally, we investigate L\'evy $\alpha$-stable symmetric jumps, with
independent displacements along $x$ and $y$ coordinates given by
$\hat{p}_x(k) = \hat{p}_y(k) = \exp(-|ak|^\mu)$, with $1 < \mu <2$.
Using Eq. (\ref{eq:hatrho_xi}), one gets thus $\hat{\rho}_\theta(k) =
\exp(-|a_\theta k|^\mu)$, with
\begin{equation}
a_\theta = a \bigl(|\cos \theta|^\mu + |\sin \theta|^\mu\bigr)^{1/\mu} .
\end{equation}
so that $\nu = 2\mu$, and $b_\theta = a_\theta^{2}/2$.  The mean
perimeter is determined by Eq. (\ref{eq:Ln_Levy2}), with
\begin{equation}
C_0 = a \, \frac{\mu \Gamma(1-1/\mu)}{\pi} \int\limits_0^{2\pi} d\theta \, \bigl(|\cos \theta|^\mu + |\sin \theta|^\mu\bigr)^{1/\mu} ,
\end{equation}
and $\gamma$ given by Eq. (\ref{eq:gamma_Levystable}) which is
independent of $\theta$.  

\begin{figure}
\begin{center}
\includegraphics[width=85mm]{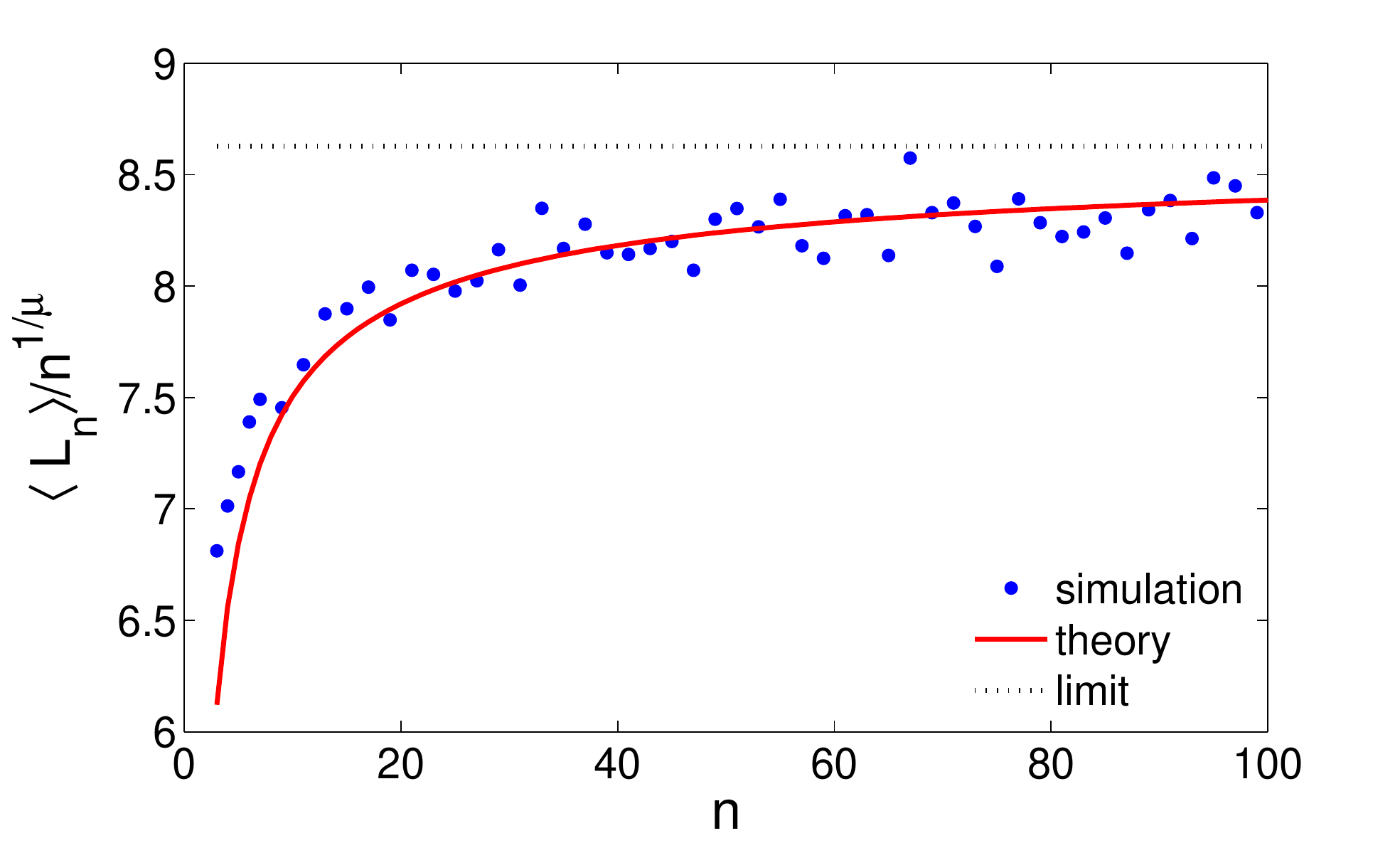} 
\end{center}
\caption{
The rescaled mean perimeter, $\langle L_n\rangle/n^{1/\mu}$, for
anisotropic planar random walk with independent L\'evy $\alpha$-stable
jump distribution with $\mu = 1.5$ and $a = 1$.  The results of Monte
Carlo simulations (shown by circles) agree well with the theoretical
prediction (\ref{eq:Ln_Levy2}) (shown by solid line). }
\label{fig:levy_stable}
\end{figure}

For $\mu = 3/2$, we obtain numerically $a^{-1} C_0 = 8.6275\ldots$ and
$a^{-1} C_1 = -5.2151\ldots$.  Figure \ref{fig:levy_stable} shows the
good agreement between the theoretical prediction (\ref{eq:Ln_Levy2})
and Monte Carlo simulations.

\section{Discussion and conclusion}
\label{sec:discussion}

To summarize, we have presented exact asymptotic results for the mean
perimeter of the convex hull of an $n$-step discrete-time random walk
in a plane, with a generic continuous jump distribution satisfying the
central symmetry assumption in Eq. (\ref{eq:p_symm}).  Explicit
results, along with simulations confirming them have been presented
for several examples of such jump distributions.  For the mean area of
the convex hull, we have derived exact results for isotropic Gaussian
jump distributions.  For jumps with a finite variance, our results
provide precise estimates of the deviations from the Brownian limit
and explain the discrepancies between the asymptotic Brownian limit
results and observed simulations for finite but large $n$.

The obtained results are particularly valuable for applications
dealing with discrete-time random processes, e.g., home range
estimation in ecology.  Given that the tracks of animal displacements
are typically recorded at discrete time steps (e.g., daily
observations) and relatively short, the subleading terms play an
important role.  The asymptotic formulas can also be used for
calibrating new estimators, based on the local convex hull, that were
proposed for the analysis of intermittent processes in microbiology
\cite{Lanoiselee17}.  Finally, the knowledge of the mean perimeter of
the convex hull can be used to estimate the scaling exponent and the
scale of symmetric L\'evy flights, for which the conventional mean and
variance estimators are useless.  

There are many interesting open problems that may possibly be
addressed using the methods presented here.  For example, the
numerical evidence suggests a possible extension of the derived
asymptotic formula for the mean area to other isotropic processes,
beyond the Gaussian case.  
Also, it would be interesting to extend our results to the case of the
convex hull of planar discrete-time random bridges (where the walker
is constrained to come back to the starting point after $n$ discrete
jumps).  For such discrete-time bridges, there are recent exact
results on the statistics of the first two maximum and the gap between
them~\cite{MMS2014} which may be useful for the convex hull problem.
One can also consider the problem with many independent discrete-time
walkers.  Finally, it would be interesting to study the statistics of
the perimeter and the area for random walks with jump distributions
that violate the reflection property in Eq. (\ref{eq:p_symm}), for
example, for walks in presence of a drift or a potential.

\section*{Acknowledgments}

DG acknowledges the support under Grant No. ANR-13-JSV5-0006-01 of the
French National Research Agency.

\appendix
\section{Asymptotic analysis}
\label{sec:Aderivation}

In this Appendix, we derive the main results (\ref{eq:Mn},
\ref{eq:Mn2}).  The derivation is based on the asymptotic analysis of
the Pollaczek-Spitzer identity (\ref{eq:PS_identity}) and extends the
earlier results from \cite{Comtet05}.  Since the function
$\phi(s,\lambda)$ from Eq. (\ref{eq:phi}) is related to the Laplace
transform of the probability density $Q'_n(z)$, it determines the
generating functions $h_m(s)$ of the moments $\langle M_n^m\rangle$
via Eq. (\ref{eq:hm}).  In turn, the asymptotic behavior of $h_m(s)$
as $s\to 1$ determines the asymptotic behavior of $\langle
M_n^m\rangle$ as $n\to \infty$.

\subsection{Derivation of generating functions}
\label{sec:Agenerating}

We recall that $M_n$ denotes the maximum of partial sums of
independent identically distributed random variables $\{\xi_k\}$:
\begin{equation}
M_n = \max\{ 0, \xi_1, \xi_1 + \xi_2, \ldots, \xi_1 + \ldots + \xi_n \} .
\end{equation}
We assume that the jump distribution is symmetric and continuous,
whereas its characteristic function $\hat{\rho}(k)$ admits the
expansion
\begin{equation}  \label{fk.1}
\hat{\rho}(k) = 1 - |ak|^{\mu} + \ldots \qquad (k\to 0) ,
\end{equation}
with an exponent $0<\mu\le 2$ and a scale $a > 0$.

First, we derive the generating function $h_1(s)$ from the
Pollaczek-Spitzer formula (\ref{eq:PS_identity}).  For this purpose,
we need to determine the expansion of this formula in powers of
$\lambda$ for small $\lambda$.  Let us first write
\begin{equation}
\phi(s,\lambda)= \exp\left[-I(s,\lambda)\right] , \quad
I(s,\lambda)= \frac{\lambda}{\pi}\int_0^{\infty}
\frac{dk}{k^2+\lambda^2}\, \ln \left(1- s\, \hat{\rho}(k)\right)\,.
\label{ISL.1}
\end{equation}
It is easy to obtain the $\lambda\to 0$ limit of $I(s,\lambda)$.  We
rescale $k= \lambda\, u$ in Eq. (\ref{ISL.1}) and take the $\lambda\to
0$ limit.  This gives, using ${\hat \rho}(k=0)=1$, a very simple
expression
\begin{equation}
I(s,0)= \frac{1}{\pi}\int_0^{\infty} \frac{du}{1+u^2}\, \ln (1- s) = \frac{1}{2}\ln (1-s)\,.
\label{IS0.1}
\end{equation}
Next we re-write
\begin{eqnarray}
\phi(s,\lambda) &=& \exp\left[-I(s,\lambda)\right] = \exp\left[-I(s,0)\right]\,\exp\left[-
\left(I(s,\lambda)-I(s,0)\right)\right]
\nonumber \\
&=& \frac{1}{\sqrt{1-s}}\exp\left[- \frac{\lambda}{\pi}\, \int_0^{\infty} \frac{dk}{\lambda^2+k^2}\, 
\ln\left(\frac{1-s\, {\hat \rho}(k)}{1-s}\right)\right]\, .
\label{phisl.1}
\end{eqnarray}
Expanding the right-hand side of Eq. (\ref{phisl.1}) up to the first
order $\lambda$ for small $\lambda$ (with $s$ fixed) gives
\begin{equation}
\phi(s,\lambda) \simeq \frac{1}{\sqrt{1-s}}\,\left[1- \frac{\lambda}{\pi} \int_0^{\infty} \frac{dk}{k^2} \ln \left(
\frac{1-s\, {\hat \rho}(k)}{1-s}\right) + o(\lambda)\right]\, .
\label{phisl.2}
\end{equation}
Note that the integral in the second term is convergent for any jump
pdf with ${\hat \rho}(k)$ satisfying Eq. (\ref{fk.1}) provided
$1<\mu\le 2$.  Taking the derivative with respect to $\lambda$,
evaluated at $\lambda = 0$, yields
\begin{equation}
h_1(s) = \sum_{n=0}^{\infty} s^n\, \langle M_n\rangle = \frac{1}{\pi (1-s)}
\int_0^{\infty} \frac{dk}{k^2} \ln \left(\frac{1-s\, {\hat \rho}(k)}{1-s}\right)\,.
\label{EMN.1}
\end{equation}
Note that this is an exact result for any jump PDF ${\hat \rho}(k)$
satisfying Eq. (\ref{fk.1}) with $1<\mu\le 2$ and it also holds for
arbitrary $s$ such that the generating function on the left-hand side
of Eq. (\ref{EMN.1}) is convergent.  In turn, when $0 < \mu \leq 1$,
the mean value $\langle M_n\rangle$ is infinite, and $h_1(s)$ is
undefined.

The second moment $\langle M_n^2\rangle$ is finite only for $\mu=2$.
Expanding Eq. (\ref{phisl.1}) up to the order $\lambda^2$ for small
$\lambda$ (with $s$ fixed) and taking the second derivative with
respect to $\lambda$ at $\lambda = 0$, one gets for any $0 \leq s < 1$
\begin{equation}
h_2(s) = \sum_{n=0}^{\infty} s^n\, \langle M_n^2\rangle = (1-s) [h_1(s)]^2 + \frac{a^2 s}{(1-s)^2} .
\end{equation}

\subsection{Jumps with a finite variance}

Here we investigate the asymptotic behavior of $h_1(s)$ as $s\to 1$
for the jump distribution with a finite variance $\sigma^2$ so that
$\hat{\rho}(k) \simeq 1 - k^2\sigma^2/2 + o(k^2)$.  First, we
represent $h_1(s)$ from Eq. (\ref{eq:h1}) as
\begin{equation}  \label{eq:h1_auxil1}
\fl
h_1(s) = \frac{1}{\pi(1-s)} \int\limits_0^\infty \frac{dk}{k^2} \left\{\ln \left(\frac{1-s(1-k^2\sigma^2/2)}{1-s}\right)
+ \ln \left(\frac{1-s\hat{\rho}(k)}{1-s(1-k^2\sigma^2/2)}\right)\right\} .
\end{equation}
The first integral can be computed explicitly, so that
\begin{equation}
h_1(s) = \frac{\sigma \sqrt{s/2}}{(1-s)^{3/2}} + \frac{I_\ve}{1-s} \, ,
\end{equation}
where $I_\ve$ denotes the second integral in Eq. (\ref{eq:h1_auxil1})
that we rewrite as
\begin{equation}
I_\ve = \frac{1}{\pi} \int\limits_0^\infty \frac{dk}{k^2} \ln \left(\frac{1 - (1-\ve^2) \hat{\rho}(k)}{\sigma^2 k^2/2 + \ve^2(1-\sigma^2 k^2/2)}\right),
\end{equation}
with $\ve = \sqrt{1-s}$.  Setting $\ve\to 0$, one gets
\begin{equation}
I_0 = \frac{1}{\pi} \int\limits_0^\infty \frac{dk}{k^2} \ln \left(\frac{1 - \hat{\rho}(k)}{\sigma^2 k^2/2}\right) = \sigma \, \gamma,
\end{equation}
with the constant $\gamma$ from Eq. (\ref{eq:gamma}), with $\sigma =
\sigma_\theta$.

Next, we consider
\begin{equation}
I_\ve - I_0 =  \frac{1}{\pi} \int\limits_0^\infty \frac{dk}{k^2} \ln \left(\frac{1 - (1-\ve^2) \hat{\rho}(k)}{\frac12 \sigma^2 k^2 + \ve^2(1- \frac12 \sigma^2 k^2)}
\, \frac{\frac12 \sigma^2 k^2}{1 - \hat{\rho}(k)} \right).
\end{equation}
Changing again the integration variable, one has 
\begin{equation}   \label{eq:auxil12}
I_\ve - I_0 =  \frac{1}{\pi \ve} \int\limits_0^\infty \frac{dk}{k^2} \ln \left(\frac{1 - (1-\ve^2) \hat{\rho}(k\ve)}
{\frac 12 \sigma^2 k^2 + (1-\frac12\sigma^2 \ve^2 k^2)} \, \frac{\frac12 \sigma^2 k^2}{1 - \hat{\rho}(k\ve)} \right).
\end{equation}
To get the next-order term, we assume the existence of the
fourth-order moment of the jumps so that
\begin{equation}  \label{eq:hatrho_Taylor}
\hat{\rho}(k\ve) \simeq 1 - \frac12 \sigma^2 k^2\ve^2 + c \sigma^4 k^4 \ve^4 + o(\ve^4),
\end{equation}
where $c = \K/24$ is related to the kurtosis $\K$ given by
Eq. (\ref{eq:a4}).  Substituting this expansion into
Eq. (\ref{eq:auxil12}), one gets
\begin{equation} \fl
I_\ve - I_0 \simeq \frac{1}{\pi \ve} \int\limits_0^\infty \frac{dk}{k^2} 
\ln \left(\frac{1 - \frac{\ve^2 k^4 c \sigma^4}{1 + \sigma^2 k^2/2}}{1 - 2k^2 \ve^2 c \sigma^2}\right) 
 \simeq \frac{1}{\pi \ve} \int\limits_0^\infty dk \frac{2c \ve^2 \sigma^2}{1 + \sigma^2 k^2/2} 
= \ve \, \sqrt{2} \, c \, \sigma .
\end{equation}
Combining these results, we conclude that
\begin{equation}  \label{eq:h1s}
\fl
h_1(s) \simeq \frac{\sigma}{\sqrt{2}\, (1-s)^{3/2}} + \frac{\sigma \gamma}{1-s} 
+ \frac{\sigma (\K/6 - 1)}{\sqrt{8}\, (1-s)^{1/2}} + o((1-s)^{-1/2}) \quad (s\to 1). 
\end{equation}
Substituting this expansion into Eq. (\ref{eq:h2}), one gets
\begin{equation}  \label{eq:h2s}
\fl
h_2(s) = \frac{\sigma^2}{(1-s)^2} + \frac{\sqrt{2}\sigma^2\gamma}{(1-s)^{3/2}} 
 + \frac{\sigma^2(\K/12 + \gamma^2 - 1)}{(1-s)} + o((1-s)^{-1})  \quad (s\to 1).
\end{equation}

In order to invert the relations (\ref{eq:h1s}, \ref{eq:h2s}), we use
the following identities:
\begin{eqnarray}
\label{eq:Tauberian1}
\sum\limits_{n=0}^\infty s^n a_n &=& \frac{1}{(1-s)^{1/2}} \,, \qquad
\sum\limits_{n=0}^\infty s^n = \frac{1}{1-s} \,, \\ 
\label{eq:Tauberian2}
\sum\limits_{n=0}^\infty s^n b_n &=& \frac{1}{(1-s)^{3/2}} \,,  \qquad
\sum\limits_{n=0}^\infty (n+1) s^n = \frac{1}{(1-s)^2} \,,
\end{eqnarray}
with
\begin{eqnarray}
a_n &=& \left(\begin{array}{c} 2n\\ n \\ \end{array}\right) 2^{-2n} \simeq \frac{1}{\sqrt{\pi n}} \biggl(1 - \frac{1}{8n} + O(n^{-2})\biggr) ,  \\
b_n &=& (2n+1) \left(\begin{array}{c} 2n\\ n \\ \end{array}\right) 2^{-2n} \simeq \frac{2\sqrt{n}}{\sqrt{\pi}} \biggl(1 + \frac{3}{8n} + O(n^{-2})\biggr) .
\end{eqnarray}
Inverting term by term, we obtain the main relations (\ref{eq:Mn},
\ref{eq:Mn2}).  Higher-order corrections can be obtained in the
same way.

\subsection{L\'evy flights}
\label{sec:ALevy}

For L\'evy flights with $\hat{\rho}(k) \simeq 1 - |a k|^\mu +
o(|k|^\mu)$, one can derive the asymptotic behavior of $h_1(s)$ as
$s\to 1$ for the case $1 < \mu < 2$.  In this case, we replace the
representation (\ref{eq:h1_auxil1}) by
\begin{equation}  \label{eq:h1_auxil2}
\fl
h_1(s) = \frac{1}{\pi(1-s)} \int\limits_0^\infty \frac{dk}{k^2} \left\{\ln \left(\frac{1-s(1-(ak)^\mu)}{1-s}\right)
+ \ln \left(\frac{1-s\hat{\rho}(k)}{1-s(1-(ak)^\mu)}\right)\right\} .
\end{equation}
As previously, the first integral can be evaluated explicitly so that
\begin{equation}  \label{eq:h1_auxil3}
h_1(s) = \frac{a \, s^{1/\mu}}{(1-s)^{1+1/\mu}} \, \frac{1}{\sin(\pi/\mu)} + \frac{I_\ve}{1-s} \,,
\end{equation}
where
\begin{equation}  \label{eq:Ive_auxil1}
I_\ve = \frac{1}{\pi} \int\limits_0^\infty \frac{dk}{k^2} \ln \left(\frac{1-(1-\ve^2)\hat{\rho}(k)}{\ve^2 + (1-\ve^2)(ak)^\mu}\right) ,
\end{equation}
with $\ve = \sqrt{1-s}$.  The first term in Eq. (\ref{eq:h1_auxil3})
provides the leading contribution.  Using the discrete form of the
Tauberian theorem,
\begin{equation}
\sum\limits_{n=0}^\infty a_n s^n = \frac{1}{(1-s)^\alpha}  \quad \Rightarrow \quad a_n \simeq \frac{n^{\alpha-1}}{\Gamma(\alpha)} \quad (n\gg 1) ,
\end{equation}
one gets the leading term to $\langle M_n\rangle$ to be $(a \mu
\Gamma(1-1/\mu)/\pi) \, n^{1/\mu}$.

In order to compute the subleading term, one needs to analyze the
integral $I_\ve$ whose behavior depends on the next-order term in the
small $k$ expansion of $\hat{\rho}(k)$ in Eq. (\ref{eq:hatrho_nu}).
Formally setting $\ve = 0$ in Eq. (\ref{eq:Ive_auxil1}), one would get
\begin{equation}  \label{eq:Ive_auxil2}
I_0 = \frac{1}{\pi} \int\limits_0^\infty \frac{dk}{k^2} \ln \left(\frac{1-\hat{\rho}(k)}{(ak)^\mu}\right) .
\end{equation}
Substituting the expansion (\ref{eq:hatrho_nu}) into this integral,
one gets the factor $\ln(1 - b k^{\nu-\mu}/a^\mu)$.  As a consequence,
the integral converges when $\nu-\mu > 1$ and diverges otherwise.
This condition naturally distinguishes two asymptotic regimes: $\nu >
\mu+1$ and $\nu \leq \mu+1$.  In the former case, the integral
converges, and the two principal terms in $h_1(s)$ are
\begin{equation}
h_1(s) \simeq \frac{a \, s^{1/\mu}}{(1-s)^{1/\mu}} \, \frac{1}{\sin(\pi/\mu)} + \frac{I_0}{1-s} + o((1-s)^{-1})  \qquad (s\to 1)\,,
\end{equation}
where $I_0$ is identical to $\gamma$ from Eq. (\ref{eq:gamma_Levy2}).
The inversion of this relation yields Eq. (\ref{eq:Mn_Levy2}).

In the case $\nu \leq \mu+1$, the subtraction of the term with
$1-|ak|^\mu$ in Eq. (\ref{eq:h1_auxil2}) was not sufficient to get the
convergent correction term.  We remedy this problem by using another
representation:
\begin{eqnarray}  \nonumber
h_1(s) = \frac{1}{\pi(1-s)} \int\limits_0^\infty \frac{dk}{k^2} && \left\{\ln \left(\frac{1-s(1-(ak)^\mu+bk^\nu)}{1-s}\right) \right.\\
\label{eq:h1_auxil4}
&& + \left. \ln \left(\frac{1-s\hat{\rho}(k)}{1-s(1-(ak)^\mu+bk^\nu)}\right)\right\} .
\end{eqnarray}
Changing the variable $u = ak (s/(1-s))^{1/\mu}$, we rewrite the first
term as
\begin{equation}
\frac{a s^{1/\mu}}{\pi(1-s)^{1+1/\mu}} \int\limits_0^\infty \frac{du}{u^2} \ln \bigl(1 + u^\mu - C u^\nu \bigr),
\end{equation}
with $C = b a^{-\nu} s^{-\nu/\mu} (1-s)^{\nu/\mu}$.  When $s\to 1$,
$C$ is small, and one can expand the logarithm to get
\begin{equation}
\frac{a s^{1/\mu}}{\pi(1-s)^{1+1/\mu}} \int\limits_0^\infty \frac{du}{u^2} \biggl\{ \ln \bigl(1 + u^\mu)  - \frac{C u^\nu}{1+u^\mu} \biggr\}.
\end{equation}
Now both integrals are convergent.  The first integral yields the same
leading term as earlier, whereas the second term provides the
subleading correction.  Finally, the second term in
Eq. (\ref{eq:h1_auxil4}) results in higher-order corrections that we
ignore.  Keeping only the leading term and the first subleading term,
we get as $s\to 1$
\begin{equation}
\fl
h_1(s) \simeq \frac{a \, s^{1/\mu}}{(1-s)^{1/\mu}} \, \frac{1}{\sin(\pi/\mu)} 
- \frac{b\, a^{1-\nu}}{(1-s)^{1+(1-\nu)/\mu}} \frac{1}{\mu \sin(\pi(\nu-1)/\mu)} \bigl(1 + o(1)\bigr).
\end{equation}
The inversion of this relation yields the large $n$ asymptotic formula
(\ref{eq:Mn_Levy1}) for $\langle M_n\rangle$.

\section{Two exactly solvable examples}
\label{sec:Aexamples}

In this Appendix, we briefly discuss two distributions, for which the
function $\phi(s,\lambda)$ can be found exactly: the symmetric
exponential distribution $\rho(z) = \frac12 e^{-|z|}$ (see
\cite{Comtet05}) and its modification $\rho(z) = \frac12 |z|
e^{-|z|}$.  These examples served us as benchmarks for checking
general results.

\subsection{Symmetric exponential distribution}

When $\rho(z) = \frac12 e^{-|z|}$, one gets $\hat{\rho}(k) =
(1+k^2)^{-1}$ and thus Eq. (\ref{eq:phi}) becomes
\begin{equation}  \label{eq:phi_model1}
\phi(s,\lambda) = \frac{1+\lambda}{\sqrt{1-s} + \lambda} \,,
\end{equation}
where we used the identity
\begin{equation}  \label{eq:auxil1}
\int\limits_0^\infty dk \frac{\ln(a^2 + b^2 k^2)}{\lambda^2+k^2} = \frac{\pi}{\lambda} \ln\bigl(a + \lambda b)
\end{equation}
to compute the integral:
\begin{equation}
\int\limits_0^\infty dk \frac{\ln(1 - s/(1+k^2))}{\lambda^2+k^2} = \frac{\pi}{\lambda} \ln \left(\frac{\sqrt{1-s} + \lambda}{1+\lambda}\right) .
\end{equation}
According to Eqs. (\ref{eq:sigma_theta}, \ref{eq:a4}, \ref{eq:gamma}),
we also obtain
\begin{equation}
\sigma^2 = 2, \qquad \K = 6, \qquad \gamma = - \frac{1}{\sqrt{2}} .
\end{equation}

The first derivative of $\phi(s,\lambda)$ from
Eq. (\ref{eq:phi_model1}) yields
\begin{equation}  \label{eq:h1_exp}
h_1(s) = \sum\limits_{n=0}^\infty s^n \langle M_n\rangle = \frac{1}{(1-s)^{3/2}} - \frac{1}{1-s} \,,
\end{equation}
from which, using Eqs. (\ref{eq:Tauberian1}, \ref{eq:Tauberian2}), one
retrieves the exact form of the first moment, which is valid for any
$n$ and was first derived in \cite{Comtet05}:
\begin{equation}  \label{eq:Mn_exp}
\langle M_n \rangle = (2n+1) \left(\begin{array}{c} 2n\\ n \\ \end{array}\right) 2^{-2n}  - 1 .
\end{equation}
Using the Stirling formula, one can get the large $n$ expansion to any
order:
\begin{equation}  \label{eq:Mn_asympt0}
\sigma^{-1} \langle M_n \rangle \simeq \frac{\sqrt{2}}{\sqrt{\pi}} \, n^{\frac 12} - \frac{1}{\sqrt{2}} + \frac{3}{4\sqrt{2\pi}}\, n^{-\frac12}
- \frac{7}{64\sqrt{2\pi}}\, n^{-\frac 32} + O(n^{-\frac52}).
\end{equation}
The first three terms agree with the general expansion (\ref{eq:Mn}).

Similarly, we get from Eq. (\ref{eq:phi_model1})
\begin{equation} \label{eq:h2_exp}
h_2(s) = \sum\limits_{n=0}^\infty s^n \langle M_n^2\rangle = \frac{2}{(1-s)^2} - \frac{2}{(1-s)^{3/2}} \, ,
\end{equation}
from which the exact formula follows using Eqs. (\ref{eq:Tauberian1},
\ref{eq:Tauberian2})
\begin{equation}  \label{eq:Mn2_exp}
\sigma^{-2} \langle M_n^2 \rangle = (n+1) - (2n+1) \left(\begin{array}{c} 2n\\ n \\ \end{array}\right) 2^{-2n}  .
\end{equation}
In the large $n$ limit, we deduce
\begin{equation}
\sigma^{-2} \langle M_n^2 \rangle \simeq n - \frac{2}{\sqrt{\pi}} \, n^{\frac 12} + 1 
- \frac{3}{4\sqrt{\pi}}\, n^{-\frac12} + \frac{7}{64\sqrt{\pi}}\, n^{-\frac 32} + O(n^{-\frac52}).
\end{equation}
The first three terms agree with the general expansion (\ref{eq:Mn2}).

Note that one can also obtain the generating function for the
cumulative distribution $Q_n(z)$.  In fact, one has for the Laplace
transform of $Q_n(z)$
\begin{equation}
\sum\limits_{n=0}^\infty s^n \L\{ Q_n(z)\}(\lambda) = \frac{1+\lambda}{\lambda \sqrt{1-s} (\sqrt{1-s}+\lambda)} \,,
\end{equation}
from which the inverse Laplace transform (with respect to $\lambda$)
yields
\begin{equation}
\sum\limits_{n=0}^\infty s^n \, Q_n(z) = \frac{1}{1-s} - \biggl(\frac{1}{1-s} - \frac{1}{\sqrt{1-s}}\biggr) e^{- z \sqrt{1-s}} \,.
\end{equation}
From this relation, one can easily get any generating function $h_m(s)$.

\subsection{Modified exponential distribution}

For the case $\rho(z) = \frac12 |z| e^{-|z|}$, one gets
\begin{equation}
\hat{\rho}(k) = \frac{1-k^2}{(1+k^2)^2} \,.
\end{equation}
Using again Eq. (\ref{eq:auxil1}), one computes the integral in
Eq.(\ref{eq:phi}) as
\begin{equation}
\int\limits_0^\infty dk \frac{\ln(1 - s\hat{\rho}(k))}{\lambda^2+k^2} = \frac{\pi}{\lambda} 
\ln \left(\frac{(\mu_+(s) + \lambda)(\mu_-(s) + \lambda)}{(1+\lambda)^2}\right),
\end{equation}
where
\begin{equation}
\mu_{\pm}(s) = \sqrt{\frac{2 + s \pm \sqrt{s^2+8s}}{2}} \,.
\end{equation}
We get thus
\begin{equation}  \label{eq:phi_model2}
\phi(s,\lambda) = \frac{(1+\lambda)^2}{(\mu_+(s) + \lambda)(\mu_-(s) + \lambda)} \,.
\end{equation}
According to Eqs. (\ref{eq:sigma_theta}, \ref{eq:a4}, \ref{eq:gamma}),
we also obtain
\begin{equation}
\sigma^2 = 6, \qquad \K = \frac{10}{3}, \qquad \gamma = \frac{1}{3\sqrt{2}} - \frac{2}{\sqrt{6}} \,.
\end{equation}

Taking the first derivative of $\phi(s,\lambda)$ from
Eq. (\ref{eq:phi_model2}) with respect to $\lambda$ yields
\begin{equation}
h_1(s) = \frac{-1}{\sqrt{1-s}} \, \frac{2\mu_- \mu_+ - (\mu_+ + \mu_-)}{\mu_+^2 \, \mu_-^2} \,.
\end{equation}
Since $\mu_- \mu_+ = \sqrt{1-s}$, $\mu_+^2 + \mu_-^2 = 2+s$, and
$\mu_+ + \mu_- = \sqrt{2+s +2\sqrt{1-s}}$, we get
\begin{equation}
h_1(s) = \frac{\sqrt{2+s +2\sqrt{1-s}}}{(1-s)^{3/2}} -\frac{2}{1-s} . 
\end{equation}
Expanding near $s\to 1$, one obtains
\begin{equation}
h_1(s) \simeq \frac{\sqrt{3}}{(1-s)^{3/2}} + \frac{\frac{1}{\sqrt{3}} - 2}{1-s} - \frac{2}{3\sqrt{3}(1-s)^{1/2}} + O(1)  ,
\end{equation}
from which
\begin{equation}
\sigma^{-1} \langle M_n\rangle \simeq \frac{\sqrt{2}}{\sqrt{\pi}}\, n^{\frac 12} + \gamma - \frac{19}{36\sqrt{2\pi}} \, n^{-\frac12} + O(n^{-\frac32}) ,
\end{equation}
in agreement with the general expansion (\ref{eq:Mn}).

Similarly, we compute $h_2(s)$ as the second derivative of
$\phi(s,\lambda)$:
\begin{eqnarray}  \nonumber
h_2(s) &=& 2 \frac{\mu_- \mu_+ + \mu_-^2 \mu_+^2 - 2\mu_- \mu_+ (\mu_- + \mu_+) + \mu_+^2 + \mu_-^2}{\sqrt{1-s}\, \mu_+^3 \, \mu_-^3}  \\
&=& \frac{6}{(1-s)^2} + \frac{2 - 4\sqrt{2+s+2\sqrt{1-s}}}{(1-s)^{3/2}} \, . 
\end{eqnarray}
As $s\to 1$, one gets
\begin{equation}
h_2(s) \simeq \frac{6}{(1-s)^2} + \frac{2-4\sqrt{3}}{(1-s)^{3/2}} - \frac{4/\sqrt{3}}{(1-s)} + O((1-s)^{-1/2}) ,
\end{equation}
from which
\begin{equation}
\sigma^{-2} \langle M_n^2 \rangle \simeq n + \gamma \frac{2}{\sqrt{\pi}} \, n^{\frac 12} 
+ 1 - \frac{2}{3\sqrt{3}} + O(n^{-\frac 12}) ,
\end{equation}
in agreement with the general expansion (\ref{eq:Mn2}).


\end{document}